%% file: main.tex
\documentclass[12pt]{article}
\input{mathsym}

\usepackage[a4paper, margin=1in]{geometry}
\usepackage{amsmath, amssymb, amsfonts}
\usepackage{graphicx}
\usepackage{hyperref}
\usepackage{xcolor}
\usepackage[normalem]{ulem}
\usepackage{tikz}
\usetikzlibrary{arrows.meta, positioning}
\usepackage{booktabs}
\usepackage{setspace}

\newcommand{\beq}{\begin{equation}}
\newcommand{\eeq}{\end{equation}}
\newcommand{\ben}{\begin{eqnarray}}
\newcommand{\een}{\end{eqnarray}}

\title{{Quantum-Gravitational Backreaction in the BTZ Background from Curved Momentum Space}}

\author{
Partha Nandi\textsuperscript{1,2}\thanks{Email: \texttt{pnandi@sun.ac.za}}, 
Mainak Roy\textsuperscript{3}, 
Langa Horoto\textsuperscript{1}, \\
Frederik G. Scholtz\textsuperscript{1}, 
Biswajit Chakraborty\textsuperscript{3} 
}

\date{
\textsuperscript{1}Department of Physics, Stellenbosch University, South Africa \\
\textsuperscript{2}National Institute for Theoretical and Computational Sciences (NITheCS), South Africa \\
\textsuperscript{3}Ramakrishna Mission Vivekananda Educational and Research Institute, Belur Math, Howrah, India
}

\begin{document}

\maketitle






\begin{abstract}

We explore how quantum properties of spacetime—specifically the curvature of momentum space—can backreact on classical gravity within a tractable semiclassical $(2+1)$-dimensional framework with negative cosmological constant. Motivated by quantum-gravity scenarios, we investigate how Planck-scale modifications of particle kinematics influence both dynamics and gravitational solutions. Starting from a first-order action, we derive an effective configuration-space description and show that particle trajectories remain geodesic, preserving the weak equivalence principle despite the underlying deformation. Coupling this modified matter sector to Einstein gravity, we obtain a deformed BTZ black hole solution. Remarkably, the local geometric structure and thermodynamic relations retain their standard form, while all quantum-gravity effects are encoded in a nonlinear mapping between the microscopic mass parameter and the ADM mass. This induces a renormalization of the horizon radius and thermodynamic quantities without altering their functional dependence. As a concrete observable consequence, we compute corrections to the return time of massless probes traveling along null geodesics between the horizon and the $AdS_3$ boundary. Our results demonstrate that Planck-scale kinematic effects can leave controlled and potentially measurable imprints on classical geometry, providing a clear and consistent bridge between quantum gravity ideas and semiclassical observables.

\end{abstract}

\section{Introduction}
We often take for granted that physics unfolds in a smooth, continuous spacetime—a backdrop against which events are localized in space and time. Yet, as local observers, what we actually measure are energies, directions, and arrival times of particles. We detect momenta—not positions. The familiar notion of spacetime arises only through a reconstruction process, built from such observational data using assumptions about symmetry, locality, and synchronization~\cite{Amelino-Camelia:2011lvm}.  {In this context, it is worth recalling that, in the spirit of noncommutative geometry, the algebra of observables can encode the underlying spacetime structure itself, generalizing the classical duality between commutative algebras and Hausdorff spaces (see, e.g., \cite{Landi,Sitarz:2014}).}\\

This realization opens the door to a radical possibility: that spacetime is not fundamental, but emergent. Instead, it may be momentum space that carries the primary geometric structure—potentially curved \cite{Kowalski-Glikman:2013rxa}, nontrivial, and even quantum in nature. In such scenarios, the geometry of momentum space itself could imprint observable effects, reshaping our understanding of locality, causality, and the very fabric of gravitational dynamics.

At the same time, reconciling general relativity with quantum mechanics remains one of the most profound challenges in theoretical physics~\cite{Rovelli:1997qj, Carlip:2001wq, Woodard:2009ns}. Gravity is geometric and deterministic, while quantum theory is algebraic and probabilistic. Near the Planck scale, these two frameworks clash, and neither can be trusted in isolation. This has motivated a wide array of quantum gravity proposals—many of which suggest that the smooth spacetime manifold breaks down at small scales, replaced by quantum or noncommutative structures~\cite{doplicher1994}.

Among bottom-up approaches to quantum gravity, one particularly fruitful line of investigation involves modifying the structure of either spacetime or momentum space. Theories based on noncommutative geometry, such as $\kappa$-Minkowski spacetime, doubly special relativity (DSR), and the principle of relative locality~\cite{Amelino-Camelia:2011lvm, Kowalski-Glikman:2013rxa, Kowalski-Glikman:2004fsz}, suggest that space and time coordinates may fail to commute or that momentum space may be curved. These features lead to deformed dispersion relations and modified symplectic structures, with potential implications even at energy scales much lower than the Planck scale.

While a consistent quantum gravity theory in \((3+1)\) dimensions remains out of reach, lower-dimensional models offer important conceptual insight. In particular, gravity in \((2+1)\) dimensions provides a remarkably rich and exactly solvable laboratory \cite{Deser:1984}. Although it lacks local gravitational degrees of freedom, \((2+1)\)-dimensional gravity exhibits global structure, allows for exact backreaction from point particles~\cite{deSousaGerbert:1990yp,tHooft:1996ziz}. Backreaction effects of quantum or noncommutative fields have also been studied in related contexts (see, e.g., \cite{Backreaction1,Backreaction2}) and admits black hole solutions such as the BTZ geometry~\cite{Banados:1992wn, carlip}. It can also be recast as a Chern--Simons gauge theory~\cite{Witten:1988hc}, highlighting deep connections between gravity, topology, and quantum algebra.

This simplified yet nontrivial framework enables a controlled exploration of how quantum-kinematic structures—like curvature in momentum space—can feed back into classical gravity, producing novel geometric and observational consequences. A related question that arises naturally is whether the masses of fundamental particles gets an upper bound of the order of Planck mass ($m_p$), if the curved momentum space owes its origin to certain Lie-algebraic type of noncommutative spacetime, as happens typically, if the deformation parameter is given by ($\frac{1}{m_p}$). In such a scenario it would be interesting to see how the different thermodynamical functions of BTZ blackhole gets modified and to what extent it can impact the evaporation process of the deformed BTZ blackhole. This is particularly interesting because the quantum  tunneling process has been shown to be slowed down in presence of noncommutativity, albeit in a different context in Moyal plane \cite{Thom:2009zz}. Moreover, the inclusion of additional topological or geometric ingredients, such as gravitational Chern-Simons terms~\cite{Deser:1982vy, Deser:1981wh} or torsional extensions in Riemann–Cartan geometry~\cite{Mielke:1991nn, Blagojevic:2003uc}, enriches the dynamics while preserving analytical control.

In this work, we explore how quantum features of spacetime—particularly the curvature of momentum space~\cite{Wagner:2021thc, Wagner:2021bqz}, inspired by noncommutative geometry—can influence classical gravity in a semiclassical $(2+1)$-dimensional setting. We begin with a noncommutative spacetime algebra in which the coordinates satisfy an $\mathfrak{su}(1,1)$ Lie algebra. Lie-algebra-type noncommutative structures and their realizations in terms of commutative coordinates and canonical momenta have been extensively studied in various dimensions, particularly in $(3+1)$ dimensions (see, e.g., \cite{Meljanac:2011,Meljanac:2008,Borowiec:2009,Kontsevich:2003,Meljanac:2017}). 

In the classical limit, this algebra gives rise to a Lie--Poisson phase space with noncommuting coordinates and commuting momenta, naturally inducing a fixed anti-de Sitter (AdS$_3$) geometry on momentum space. This framework highlights a duality between noncommutative spacetime and curved momentum space, reflecting Max Born’s principle of phase space reciprocity~\cite{born1938}, which posits a fundamental symmetry under the exchange $\hat{x} \leftrightarrow \hat{p}$ as a guiding principle toward unifying quantum theory and gravity. This perspective has been further explored in recent work by Nandi et al.~\cite{Nandi:2023tfq}, where the emergence of Lorentz covariance from quantum dynamics was investigated.

While Born's reciprocity is a central inspiration, the notion of curved momentum space has even deeper roots. It can be traced back to Riemann’s 1854 work on the geometry of abstract manifolds and was further developed by Finsler and Cartan into generalized geometric frameworks now known as Hamiltonian and Lagrangian geometry~\cite{Finsler1918,Cartan1934}. These mathematical ideas reentered physics through the works of Gol’fand and Tamm~\cite{Golfand1959,Golfand1962,Golfand1963,Tamm1965,Tamm1972}, who investigated quantum theories with curved momentum space \cite{Franchino-Vinas:2023rcc}, and ultimately led to the development of quantum groups and noncommutative geometry~\cite{Lizzi:2020tci}.\\

More recently, these ideas have coalesced in the framework of \emph{Born geometry}~\cite{Freidel2019}, which unifies symplectic, complex, and metric structures in phase space, providing a geometric foundation for both quantum theory and general relativity. Within this lineage, our model can be seen as a concrete realization of Majid’s proposal of \emph{co-gravity}~\cite{majid2015}, where curvature can reside in momentum space as a dual manifestation of quantum spacetime structure. Here, the AdS$_3$ momentum space not only encodes the dual to spacetime noncommutativity but also modifies classical gravitational dynamics, offering a novel semiclassical bridge between quantum geometry and curved spacetime.\\

In our setting, the AdS geometry of momentum space is not dynamical; it is fixed by the underlying noncommutative algebra. We interpret this curvature as a semiclassical imprint of quantum geometry and investigate its consequences for classical gravitational observables. Starting from a first-order formalism based on a fuzzy $\mathbb{R}^{1,2}_\star$ spacetime, we derive an effective configuration-space action for a relativistic, spinless point particle. {The resulting dynamics exhibits mass-dependent kinematical modifications; however, the spacetime trajectories remain geodesic, ensuring that the weak equivalence principle is preserved (see appendix E).} Notably, in $(2+1)$ dimensions, the Planck mass and Planck length scale differently, allowing Planck-scale effects to be captured already in the classical limit $\hbar \to 0$, without requiring $G \to 0$ simultaneously, while keeping the ratio $\hbar/G$ fixed (see Appendix~A).

The corresponding energy-momentum tensor is then computed and used as a source in the semiclassical Einstein equations. Solving these yields a deformed BTZ black hole geometry \cite{carlip}, whose ADM mass, Hawking temperature, and entropy all receive corrections determined by the momentum space curvature and becomes bounded. We also study the impact of these corrections on the semi-classical emission process of BTZ blackholes associated with deformed ADM mass parameter. For this we consider, the dual effects of return time of a massless quanta , as it travels along a null geodesic back and forth, from near the horizon to $AdS_3$ boundary, stemming solely, from the curved nature of momentum space and reduction of emission mass after emission. These results illustrate how quantum spacetime features—encoded in the momentum space structure—can manifest as observable modifications in classical gravitational backgrounds.

\begin{figure}[h]
\centering
\begin{tikzpicture}[
  every node/.style={transform shape},
  box/.style = {
    draw,
    rounded corners,
    minimum width=3.2cm,
    minimum height=0.8cm,
    align=center,
    font=\footnotesize
  },
  ->, >={Stealth[length=3pt,width=4pt]}
]

\node[box] (A) at (90:4.8cm) {Quantum Spacetime\\(Noncommutative $\mathbb{R}^{1,2}_{\star}$)};
\node[box] (B) at (45:4cm) {Lie--Poisson Algebra\\$\{x^a, x^b\} = \dfrac{1}{m_p} \epsilon^{abc} x_c$};
\node[box] (C) at (0:4cm) {Curved Momentum Space\\( AdS geometry )};
\node[box] (D) at (-45:4cm) {Canonical Coordinates $q^\mu$\\via Bopp shift};
\node[box] (F) at (-90:4cm) {Modified Stress Tensor\\$T^{\mu\nu}(q)$};
\node[box] (G) at (-135:4cm) {Einstein Equations\\$G_{\mu\nu} + \Lambda g_{\mu\nu} = 8\pi G T_{\mu\nu}$};
\node[box] (H) at (180:4cm) {Deformed BTZ Geometry};
\node[box] (I) at (135:4cm) {Thermodynamics\\$M_{\text{ADM}}, T_H, S$};
\node[box] (E) at (0,0) {Effective Action\\$S_{\text{eff}}[q, g_{\mu\nu}]$};

\draw[->] (A) -- (B);
\draw[->] (B) -- (C);
\draw[->] (C) -- (D);

\draw[->, bend right=20] (D) to (F);

\draw[->, bend right=20] (F) to (G);

\draw[->] (G) -- (H);
\draw[->] (H) -- (I);

\draw[->] (B) -- (E);
\draw[->] (D) -- (E);
\draw[->] (E) -- (F);

\end{tikzpicture}
\caption{Compact flowchart showing the progression from non commutative spacetime to curved momentum space, deformed particle dynamics, and back reacted geometry with modified black hole thermodynamics. }
 
\label{fig:flowchart_circular}
\end{figure}

The paper is organized as follows. In Section~2, we introduce the noncommutative \((2+1)\)-dimensional model in which the spacetime coordinates satisfy an $\mathfrak{su}(1,1)$ algebra. By augmenting these with commuting momentum operators, we construct a deformed phase space whose classical limit exhibits a fixed AdS$_3$ geometry on momentum space. In Section~3, we formulate the action for a relativistic, massive, spinless particle. This action respects both the Poincaré symmetry $\mathcal{ISO}(2,1)$ and the diffeomorphism invariance of the curved momentum space. We relate the geodesic distance on this momentum space to a deformed dispersion relation, which allows us to identify the observable (renormalized) mass of the particle, and show that it is bounded above.

In Section~4, we derive the effective configuration-space action describing the particle trajectory and compute the corresponding energy-momentum tensor. Section~5 is devoted to solving the semiclassical Einstein equations with this modified stress tensor as a source, resulting in a deformed BTZ black hole geometry. The associated thermodynamic quantities—including the ADM mass, Hawking temperature, and entropy—are computed and found to carry explicit corrections determined by the curvature of momentum space. More specifically, we study the impact of curved momentum space on the return time of travel of a massless particle, from near the horizon to $AdS_3$ boundary (as illustrated schematically in Fig.~\ref{fig:flowchart_circular}). Finally, in Section~6, we summarize our results and outline possible future directions. 

Seven appendices are included to provide detailed derivations, technical clarifications, and supplementary background material.

\section{Emergent Curved Momentum Space with Minimal Deformation}
Let us first of all consider the noncommutative fuzzy $\mathbb{R}^{(1,2)}_{\star}$ with Lorentzian signature

\begin{eqnarray}
[\hat{x}^a, \hat{x}^b] &=& i\, \epsilon^{ab}\!{}_c \hat{x}^c\label{100}
\end{eqnarray}

Here the  of operator-valued spacetime  coordinates $\hat{x}^{a}$ of fuzzy $\mathbb{R}^{(1,2)}_{\star}$ taken to be dimensionless and  fulfill the above $\mathfrak{su(1,1)}$ algebra \cite{Freidel:2006}.This structure of the commutator algebra$(1)$ remains stable if these operator valued coordinates $\hat{x^a}$'s are subjected to $(2+1)D$ Lorentz transformation $\mathcal{SO}(1,2):\hat{x^a}\rightarrow \hat{x'}^{a} = \Lambda^{a}{}_{b} \hat{x}^b
$ with $\Lambda \in \mathcal{SO}(1,2)$. Note that we have adopted the mostly positive convention $(-,+,+)$ for the signature here for the metric $\eta_{ab}$ and the reference sequence $\epc{0}{1}{2}=1=-\epsilon_{012}$ for the alternating Levi-Civita symbol. \footnote{We employ the identity \[
\varepsilon^{ijk} \varepsilon^{lmn} = - 
\begin{vmatrix}
\eta^{il} & \eta^{im} & \eta^{in} \\
\eta^{jl} & \eta^{jm} & \eta^{jn} \\
\eta^{kl} & \eta^{km} & \eta^{kn}
\end{vmatrix}
\]
 and other identities which are obtained by contractions with $\eta_{il}$ etc.}\\

 It is evident from the commutation relation (\ref{100}) that constant infinitesimal spacetime translations of the form \( \delta \hat{x}^a = \xi^a \) cannot be generated by any translation generators via the relation  
\begin{equation}
\delta \hat{x}^a = i \xi^b [\hat{p}_b, \hat{x}^a],\label{101}
\end{equation}
particularly when the momentum generators commute, i.e.,  
\begin{equation}
[\hat{p}_a, \hat{p}_b] = 0.\label{102}
\end{equation}


Our goal is to identify a deformation of the standard canonical Heisenberg algebra that remains consistent with the noncommutative spacetime structure~\eqref{100}, by enforcing the Jacobi identities involving the triplets \( (\hat{x},\hat{x},\hat{p}) \), \( (\hat{p},\hat{p},\hat{x}) \), and their cyclic permutations.

As a trial solution for the infinitesimal deformed translation rules, we adopt the following ansatz, inspired by analogous constructions in flat Snyder and Snyder--de Sitter spacetimes developed by Banerjee \textit{et al.}~\cite{Banerjee:2006wf,Banerjee:2011ag}:
\begin{equation}
\delta \hat{x}^a = \xi^a + \alpha (\xi \cdot \hat{p})\, \hat{p}^a + \beta\, \epsilon^{abc} \hat{p}_c \xi_b,\label{103}
\end{equation}
where \( \alpha \) and \( \beta \) are coefficients to be determined by demanding compatibility with the underlying noncommutative structure.

In particular, we require that these modified transformations preserve the \(\mathfrak{su}(1,1)\) algebra satisfied by the coordinates. Enforcing the Jacobi identities leads to a unique determination of the coefficients:
\begin{equation}
\alpha = -\frac{1}{4}, \qquad \beta = \frac{1}{2}.\label{104}
\end{equation}

The resulting deformed Heisenberg algebra takes the form
\begin{equation}
[\hat{x}^a, \hat{p}_b] = i {E^{-1}(p)^a}_b, \label{3}
\end{equation}
where the matrix \( E^{-1}(p) \) is defined as
\begin{equation}
{(E^{-1}(p))^a}_b = \delta^{a}_{~b} -\frac{1}{4} \hat{p}^a \hat{p}_b + \frac{1}{2}\, \epsilon^a{}_{bc} \hat{p}^c. \label{4}
\end{equation}

It may be noted that we have raised/lowered the Lorentzian indices $a,b,\dots$ occurring in both $\hat{x}$'s and $\hat{p}$'s, taking values $\{0,1,2\}$, using the metric $\eta^{ab}$ / $\eta_{ab}$, implying we introduced an orthonormal basis. We now want to demonstrate here, however, that this non-trivial structure of the matrix $E^{-1}(p)$, which is entirely momentum dependent, owes its origin to the emergent curved nature of the momentum space in the classical $(\hbar \rightarrow 0)$ limit of this toy model of quantum spacetime \eqref{100} in the vicinity of the Planck scale, so that the contravariant components of momentum $p^{a}$ can still be used to ``coordinatize" the curved momentum space. In other words, the flat momentum space can be thought of as providing a coordinate chart for the curved momentum space—at least patchwise.

In contrast, the coordinate space remains flat even in this regime of quantum gravity \cite{Girelli:2004md}, which can be captured entirely by the classical limit    $(\hcut \rightarrow 0)$ itself in our $(2+1)D$ model. In this context we would  like to mention that this scenario is a bit different from the corresponding $(3+1)$ dimensional case \cite{AmelinoCamelia:2003xp}. Here in this $(2+1)D$ case we don't need to take the $G\rightarrow 0$ simultaneously   holding their ratio $\frac{G}{\hcut}$  fixed; only a classical limit will suffice (see Appendix A). In fact, in this $(2+1)D$,    the Planck length and mass  scale are given by $l_{p} \sim \hbar G$ and,  $m_p \sim \frac{1}{G}$ respectively. To see it more explicitly, let's introduce a length scale $2\lambda$ and Planck's constant $\hbar$ by scaling now the dimension full $\hat{x^a}$'s and $\hat{p}^a$'s as $\hat{x}^a \rightarrow \frac{\hat{x}^a}{2\lambda}$ and $\hat{p}^a \rightarrow \frac{2\lambda\hat{p}^a}{\hbar}$ in the above set of equations \eqref{100} to \eqref{103} to get, 

\begin{eqnarray}
   \left[ \hat{x}^a, \hat{x}^b \right] &=& 2i\lambda \epsilon^{abc} \hat{x}_c \\\label{5}
   \left[ \hat{p}_a, \hat{p}_b \right] &=& 0 \\\label{106}
   \left[ \hat{x}^a, \hat{p}_b \right] &=& i \hbar \vier{a}{b}{p}\label{7}
   \end{eqnarray}

where
\begin{eqnarray}
\trib{a}{b}(p) &=& \tribform{a}{b}{c}{\frac{2\lambda}{\hbar}}{\frac{4\lambda^2}{\hbar^2}}
\label{111}
\end{eqnarray}

Now we can identify the length scale given by the parameter $\lambda$ occurring in \eqref{104} relates to the Planck length scale $l_p$: $\lambda = \hbar G$. we see that although all the phase space variable commutes among themselves in the classical limit$(\hbar \rightarrow 0)$ i.e.:
\begin{eqnarray}
\com{x^a}{x^b}=0=\com{x^a}{p_b},
\label{9}
\end{eqnarray}
the symplectic structures gets deformed and this is obtained by applying the standard rule to any pair of phase space variables $A,B$ as:
\begin{eqnarray}
\left\{ \hat{A}, \hat{B} \right\}&=&\lim_{\hbar \to 0}\frac{\com{\hat{A}}{\hat{B}}}{i\hbar}
\label{113}
\end{eqnarray}
to get,
\begin{eqnarray}
\left\{ x_a, x_b \right\}=\frac{1}{m_p}\epsilon_{ab}^{~~c}x_c;\pb{p_a}{p_b}=0;\pb{x_a}{p^b}=\exviecm{p}{a}{b}
\label{114}
\end{eqnarray}
where,
\begin{eqnarray}
\exvielm{p}{a}{b}&:=&\ikd{a}{b}+\frac{1}{m_p}\eps{a}{b}{c}p^c-\dfrac{1}{{m_p}^2}p_ap^b; ~~~~    \frac{1}{m_p}=\frac{2\lambda}{\hbar}
\label{12}
\end{eqnarray}
Note that here the deformation parameter is $\frac{1}{m_p}$ and the undeformed structure is recovered only in the limit $m_p \rightarrow \infty$. This limit will be referred to as the commutative limit in the sequel. 

So far, we have treated the Latin indices, $a,b,...$ etc, to be the Lorentz indices, where the indices are lowered/raised using the flat metric $\eta_{ab}$ for both $x^a$'s and $p^a$'s and we write $x_a=\eta_{ab}x^b$ and $p^a=\eta^{ab}p_b$.  
Now it is trivial to see that these $x_a$'s can be represented by a vector field on our flat momentum space, $ \mathcal{P}_{\infty} $ as, 
\begin{eqnarray}
\rho(x_a)=\exviecm{p}{a}{b}\pd{}{p^b}
\label{13}
\end{eqnarray}
which acts by definition,on any arbitary function $F(p)$ of momentum adjointly as, 
\begin{equation}
    \rho(x_a) \triangleright F(p):=\left[\exviecm{p}{a}{b}\pd{}{p^b},F(p)\right]
\end{equation}
so that their simple commutator algebra can be shown to satisfy \eqref{111} entirely, upto isomorphism.

\begin{equation}
    \left[ \rho(x_a), \rho(x_b) \right]=\frac{1}{m_p}\epsilon_{ab}^{~~c}\rho(x_c);~ \left[ \rho(x_a),p^b\right]=(E^{-1}(p))_a{}^b \ .
\end{equation}

Here, $\rho(x_a)$'s can be though of appropriate Hamiltonian vector fields $X_{x^a}$ used in geometric quantization. In fact, our construction of $\rho(x_a)$  here was inspired by this geometric quantization. We have provided a brief discussion of this geometric quantization in Appendix B. 
It is worth recalling that we have already adopted the convention that Lorentz indices
$a,b,\dots$ are raised and lowered using the flat metric $\eta_{ab}$. This ensures that the $x_a$ transform covariantly under
$SO(1,2)$, while $x^a$ transform contravariantly. Since the metric tensor is a bilinear map on the tangent space,
\begin{eqnarray}
g: T_Q(\mathcal{P}_\infty)\times T_Q(\mathcal{P}_\infty) \;\rightarrow\; \mathbb{R},
\end{eqnarray}
Lorentz covariance requires that its components in the $\{x_a\}$ basis be the invariant tensor
$\eta_{ab}$, because the Lorentz matrices preserve $\eta_{ab}$ via
\begin{eqnarray}
{\Lambda_a}^c \, {\Lambda_b}^d \, \eta_{cd} = \eta_{ab}.
\end{eqnarray}
Hence we must have
\begin{eqnarray}
g(x_a, x_b) &=& \eta_{ab},
\label{109}
\end{eqnarray}
which shows that the $\{x_a\}$ furnish a non-holonomic but orthonormal basis on the tangent
space of the momentum manifold.

\medskip

The absence of any inherent length scale at the classical level of the emergent commutative
spacetime $\mathcal{M}$ \eqref{109} further supports this conclusion, allowing us to treat
$\mathcal{M}$ as effectively flat. This flatness is also reflected at the quantum
level, since the momentum operators obey a commutative algebra, consistent with an underlying
flat spacetime structure in the commutative limit. Accordingly, the $x_a$ may be regarded as
coordinate-like Killing vectors on $\mathcal{P}_\infty$. However, the converse is not true: the flat
form of $g(x_a,x_b)$ does not by itself imply that the full momentum space is flat, unless the
$\{x_a\}$ form a holonomic basis, which occurs only in the limit $m_p \to \infty$.

\medskip

However, this result would be incompatible with a genuinely flat momentum space $\mathcal{P}_\infty$,
since setting $g\!\left(\frac{\partial}{\partial p^a}, \frac{\partial}{\partial p^b}\right) = \eta_{ab}$ would
require the inverse frame field $E^{-1}(p)$ to be a Lorentz transformation,
$E^{-1}(p) \in \mathcal{SO}(1,2)$, which is clearly not the case. This indicates that the assumption of
a globally flat momentum space must be relaxed: $\mathcal{P}_\infty$ should be replaced by a curved
momentum space $\mathcal{P}_{m_p}$ that depends on the finite mass scale $m_p$, and reduces to
$\mathcal{P}_\infty$ only in the limit $m_p \to \infty$. Accordingly, Eq.~\eqref{114} should be understood
as holding point wise in the tangent spaces of the curved momentum manifold.

\medskip

At this stage, it is convenient to make use of Greek indices like $\mu,\nu,\dots \in \{0,1,2\}$ as superscripts in $p^{\mu}$ to denote the momentum variables required to coordinatize $\mathcal{P}_{m_p}$. In contrast, Latin indices like $a,b,\dots$ will be used to denote components of vector/tensor fields in an \textit{orthonormal} basis of tangent space $T_Q(\mathcal{P}_{m_p})$ or its dual cotangent space $T^*_Q(\mathcal{P}_{m_p})$. With this, we can rewrite Equation \eqref{113} by suppressing the representation index $\rho$ as, 

\begin{eqnarray}
x_a = \left(E^{-1}(p)\right)_a{}^{\mu} \frac{\partial}{\partial p^{\mu}}
\label{15}
\end{eqnarray}

where $\left(E^{-1}\right)_a{}^{\mu}$ has essentially the same structure as that of $\left(E^{-1}\right)_a{}^{b}$; one just requires to replace the Local Lorentz index $b$ with the world index $\mu$ i.e.$b\rightarrow \mu$.\footnote{In the convention that we have adopted $p^{\mu}$ is identical to $p^a$. But $p^{\mu}$ ceases to be a 3 vector under the diffeomorphism in $\mathcal{P}_{m_p}$,although it retains its property as Lorentz 3-vector. Consequently, raising/lowering of indices in $p^{\mu}$ can only be done by using the flat metric $\eta^{\mu\nu}/\eta_{\mu\nu}$  to write $p_{\mu}=\eta_{\mu\nu}p^{\nu}$. Thus $p^2 = \eta_{\mu\nu} \, p^\mu p^\nu$ is an $\mathcal{SO}(1,2)$ scalar and one can write $\frac{\partial p^2}{\partial p^{\lambda}} = 2p_{\lambda}$. We will identify the momentum components \( p^a \) explicitly in the sequel.}  In fact,   the matrix $\left(E^{-1}\right)_a{}^{\mu}$ can be regarded as the triad (i.e. the counterparts of tetrads/vielbeins in (3+1)D)- a $(3\times 3)$ matrices relating the orthonormal but non-holonomic basis like $\{ x_a \}$ to the non orthonormal but holonomic basis $\{\pd{}{p^{\mu}}\}$ of the tangent space,
\begin{eqnarray}
T_Q({\mathcal{P}_{m_p}}) = \text{Span} \{ x_a \}=\text{Span}\{\pd{}{p^{\mu}}\}
\label{16} 
\end{eqnarray}
Now,using\eqref{114}  we get,
\begin{eqnarray}
\eta_{ab}=g(x_a,x_b)={(E^{-1})_a}^{\mu}(p){(E^{-1})_b}^{\nu}(p)g_{\mu\nu}(p)
\label{17}  
\end{eqnarray}
\begin{eqnarray}
g_{\mu\nu}(p)=g\left(\pd{}{p^{\mu}},\pd{}{p^{\nu}}\right)
\label{18}
\end{eqnarray}
Inverting one gets, 

\begin{eqnarray}
g_{\mu\nu}={(E)^a}_{\mu}{(E)^b}_{\nu}\eta_{ab} \quad \text{and} \quad g^{\mu\nu}=\eta^{ab}{(E^{-1})_a}^{\mu}{(E^{-1})_b}^{\nu} 
\label{19}
\end{eqnarray}

Here $E(p)^a{}_{\mu}$ is the inverse of the matrix $\left(E^{-1}(p)\right)_b{}^{\nu}$ fulfilling,

\begin{eqnarray}
{(E)^a}_{\lambda}{(E^{-1})_{a}}^{\mu}=\kd{\lambda}{\mu} \quad \text{and} \quad {(E)^a}_{\lambda}{(E^{-1})_{b}}^{\lambda}=\ikd{a}{b} 
\label{20}
\end{eqnarray}  
\footnote{Note that, $E^a{}_{\mu}$ is a composite object,unlike $p^{a}$ and has a Lorentz index $a$, and world index $\mu$ and is obtained from that of $E_a{}^{\mu}$ by simultaneously raising $a$ and lowering $\mu$ by using $\eta^{ab}$ and $g_{\mu\nu}$ respectively as, $E(p)^a{}_{\mu}=\eta^{ab}g_{\mu\nu}E^{-1}(p)_b{}^{\nu}$}
Finally substituting the expression of ${E^{a}}_{\mu}$, one can compute the metric $g_{\mu\nu}$ and its inverse $g^{\mu\nu}$ as,
\begin{eqnarray}
g_{\mu\nu}(p) = \dfrac{m_p^2}{m_p^2 - p^2} \left[ \eta_{\mu\nu} + \dfrac{p_{\mu} p_{\nu}}{m_p^2 - p^2} \right]
\label{121}
\end{eqnarray}

and 

\begin{eqnarray}
g^{\mu\nu}(p) = \left(1 - \frac{p^2}{m_p^2}\right) \left[\eta^{\mu\nu} - \frac{p^{\mu} p^{\nu}}{m_p^2}\right]
\label{22}
\end{eqnarray} 

Interestingly,the metric \eqref{121} is conformal to de-Sitter metric ($dS_{3}$) with a conformal factor $\dfrac{m^2_p}{m^2_p-p^2}$. To see this, consider the flat $(1+3)D$ flat Minkowski space with line element
\begin{eqnarray}
 dS^2=G_{MN}dp^Mdp^N;M,N\in [0,1,2,3]
 \label{23}
 \end{eqnarray} 
and $G_{MN}=diag(-1,1,1,1)$.\footnote{The (1+3)-dimensional flat momentum space can be viewed as an ambient space in which the deformed manifold \( \mathcal{P}_\infty \) is embedded. In this setting, the first three components of the higher-dimensional momentum vector \( p^M \) correspond to \( p^a \), such that
$p^M = (p^a, p^3)$.} We can now obtain the induced metric $\widetilde{g_{\mu\nu}(p)}$ on the hypersurface $\mathcal{S}$, defined by,
\begin{eqnarray}
G_{MN}p^{M}p^{N}=m_{p}^{2}
\label{24} 
\end{eqnarray}
 by eliminating $p^3$ to get
\begin{eqnarray}\label{25}
\widetilde{g_{\mu\nu}(p)} = \eta_{\mu\nu} + \frac{p_{\mu}p_{\nu}}{m_p^2-p^2}
\end{eqnarray}

 which reproduces the metric $g_{\mu\nu}(p)$ \eqref{121}, up to the above-mentioned conformal factor. The associated Ricci tensor and curvature scalar of this de-Sitter metric (30)  is well known and are given by 
 \begin{eqnarray}\label{X}
   \tilde{R}_{\mu\nu}=\frac{2}{m^{2}_{p}}\tilde{g}_{\mu\nu},~~  \tilde{R}=\frac{6}{m_p^2}
 \end{eqnarray}

Since, this is one of three maximally symmetric spaces in (1+2)D (flat, $dS_3$, $AdS_3$) it becomes interesting to study the nature of the manifold described by the entire metric \eqref{121}. It turns out, however, the entire metric (\ref{121}) with the inclusion of the conformal factor switches to $AdS_{3}$. This can be seen easily by starting with the line element for the entire momentum space, which can be written as

\begin{equation}
    ds^2 = g_{\mu\nu}(p) \, dp^{\mu} dp^{\nu} = \Omega^2(p) \, \widetilde{g_{\mu\nu}(p)} \, dp^{\mu} dp^{\nu}
\end{equation}

where the  conformal factor is now given by:
\begin{equation} \label{confo}
    \Omega(p) = \dfrac{1}{\sqrt{1 - m_{p}^{-2} p^2}}.
\end{equation}

The corresponding Ricci tensor $R_{\mu\nu}$ and the curvature scalar $R$ can now be computed easily to find (see Appendix~B1):
\begin{align} \label{ricci}
    R_{\mu\nu} &= -\frac{2}{m_p^2}g_{\mu\nu} \\
    R &= \tilde{R} \left(1 - m_{p}^{-2}p^2\right) + 6m_{p}^{-4}p^2 - 12m_{p}^{-2}=-\frac{6}{m_p^2}
\end{align}

where we have made use of \eqref{X}. This shows that,this is an Einstein manifold with constant negative curvature and therefore  $g_{\mu\nu}(p) $  (\ref{121}) indeed corresponds to the AdS metric (at least locally: $\mathcal{P}_{m_{p}}=AdS_{3})$ which clearly becomes flat ($\mathcal{P_{\infty}}$) with $R_{\mu\nu}=0=R$ in the commutative $m_p\rightarrow \infty$ limit. We have also provided an alternative method of computation for $R_{\mu\nu}$ and $R$ in Appendix B2.

In a certain sense, the momenta \( p^a \) provide a direct coordinatization of the \( dS_3 \) momentum space and an indirect one for \( AdS_3 \), where the nontrivial conformal factor plays a crucial role. To clarify this analogy, recall that a direct coordinatization of \( AdS_3 \) is typically obtained by embedding the \( AdS_3 \) hyperboloid into a flat \((2+2)\)-dimensional momentum space.

Before concluding this section, we emphasize that the phase space structure in equation~\eqref{114} is by no means unique. In fact, a simple but nonlinear momentum-space coordinate transformation of the form
\begin{equation}
    p^{\mu} \rightarrow p^{'\mu} = \frac{p^{\mu}}{\Omega(p)}
\end{equation}
can yield a structure similar to that proposed in~\cite{Freidel:2006}, where the noncommutative brackets take the form
\[
\{ x_a, p^{'\mu} \} = \sqrt{1 - \frac{p'^2}{m_p^2}}\, \delta_a^{~\mu} +\frac{1}{m_{p}} \epsilon_{a}^{~\mu\nu} p'_\nu.
\]
However, in this case, the factor \( \frac{1}{\Omega(p)} \) enters as a prefactor to the Kronecker delta and, when expanded, generates an infinite series in powers of \( p^2 \). By contrast, our phase space structure terminates at quadratic order in momenta, thus representing a minimal deformation of the standard commutative algebra~\cite{Koch:2004ud}.

Moreover, our choice of momentum coordinates \( p^{\mu} \)—identified with the first three Cartesian components \( (p^0, p^1, p^2) \) of the ambient flat \((1+3)\)-dimensional momentum space—possesses a natural interpretation, up to Lorentz transformations in \( SO(1,2) \).



\section{Action of a relativistic spin-less but massive point particle and Dispersion Relation}

Let us consider the following first-order form of the Lagrangian
describing the dynamics of a single spin-less massive point particle
of mass $m$ propagating in a classical commutative spacetime:
\begin{eqnarray}\label{260}
L &=& -q_{\mu} \dot{p}^{\mu} - \Lambda(f(p^2) + M^2),
\qquad
\dot{p}^{\mu} = \frac{dp^{\mu}}{d\tau}.
\end{eqnarray}

Here $\tau$ parametrizes the worldline of the particle and
$\Lambda$ is a Lagrange multiplier enforcing the constraint
\begin{eqnarray}\label{274}
 f(p^2)+M^{2}\approx 0 ,
\end{eqnarray}
where $f$ is an invertible function whose explicit form will be
determined below.

A direct inspection of Eq.~\eqref{260} allows one to read off the
symplectic structure of the theory,
\begin{equation}\label{141}
\{ q_{\mu}, q_{\nu} \} = 0, \qquad
\{ p_{\mu}, p_{\nu} \} = 0, \qquad
\{ q_{\mu}, p^{\nu} \} = \delta_{\mu}^{\ \nu}.
\end{equation}
These brackets can be interpreted either as Dirac brackets arising
from the second-class constraints of the first-order Lagrangian,
or equivalently as the symplectic brackets obtained in the
Faddeev--Jackiw formulation~\cite{faddeev1988}.  The constraint
\eqref{274} retains the status of a first-class constraint.

{Before specifying the relation between the variables $q_\mu$ and the
noncommutative generators $x^a$, it is useful to clarify the geometric
role of the latter. The variables $x^a$ satisfy the Lie--Poisson
algebra (\ref{114}) and should therefore be interpreted as generators
of translations on momentum space rather than as spacetime coordinates. Because of this
nontrivial Poisson structure they form a nonholonomic frame on the
momentum--space manifold and can be interpreted as labeling the symplectic leaves of the phase space.
Consequently the variables $x^a$ do not furnish a commuting coordinate
system and cannot be used directly to parametrize a $(2+1)$-dimensional
pseudo--Riemannian spacetime manifold.}

{To obtain a configuration--space description suitable for coupling to
classical gravity we therefore introduce the variables
\begin{eqnarray}\label{142}
q_{\mu}:=x_a{(E(p))^a}_{\mu},
\end{eqnarray}
which will play the role of configuration-space coordinates.
Using Eq.~\eqref{17}, one can verify that these variables furnish canonical coordinates satisfying the symplectic structure \eqref{11}.}

{In particular, the canonical coordinate $q_\mu$, defined through
its Poisson bracket $\{q_\mu,p^\nu\}=\delta_\mu^{\ \nu}$,
generates the Hamiltonian vector field
\begin{equation}\label{30}
\{q_\mu,\cdot\}=\frac{\partial}{\partial p^\mu},
\end{equation}
 which corresponds to a holonomic (i.e., coordinate-induced) but generally non-orthonormal basis for vector fields on momentum space.}

{From this perspective the variables $q^\mu$ provide a holonomic coordinate chart for the configuration space associated with the phase-space description. The relation \eqref{28} should
therefore be understood as expressing the noncommutative generators
$x^a$ in terms of this coordinate basis through the
momentum--dependent triad $(E(p))^a{}_{\mu}$.  In this way the
momentum dependence resides in the frame transformation, while the
coordinates $q^\mu$ themselves furnish a consistent chart for
describing particle trajectories.}


We can now make use of  these coordinates $q^{\mu}$   to define the Lorentz   $(\mathfrak{so(1,2)})$ generators as, 
\begin{eqnarray}\label{31}
M_{{\mu}{\nu}}&:=&q_{\mu}p_{\nu}-q_{\nu}p_{\mu}
\end{eqnarray} 
fulfilling the entire Poincare $\mathfrak{iso}(1,2)$  algebra:
\begin{eqnarray}\label{32}
\left[ M^{\mu\nu}, M^{\rho\sigma} \right] &=& i \left( \eta^{\mu\rho} M^{\nu\sigma} - \eta^{\mu\sigma} M^{\nu\rho} - \eta^{\nu\rho} M^{\mu\sigma} + \eta^{\nu\sigma} M^{\mu\rho} \right)\\
\left[P^\mu, M^{\nu\lambda}\right] &=& i \left( \eta^{\mu\nu} P^\lambda - \eta^{\mu\lambda} P^\nu \right)
\end{eqnarray}

 {Clearly, this furnishes us with two Casimirs. One of which is the Pauli-Lubanski scalar $W = \frac{1}{2} \epsilon^{\mu\nu\rho} P_\mu M_{\nu\rho}$ gives the spin component as $W=ms$ and the other $p_{\mu}p^{\mu}$ the mass content as $p_{\mu}p^{\mu}=-m^2$ for our $p^{\mu}$,taken to be time-like.Since, we are dealing with spin-less particle, the only Casimir $p_{\mu}p^{\mu}$ is of relevance, as $W=0$ in our case. It immediately, follows that the deformed dispersion relation will involve a deformation of this Casimir operator, but still remaining invariant under $\mathcal{ISO}(1,2)$ transformation, and therefore the deformation of $p^2$ should be of the form of $f(p^2)$ so that the action \( S = \int d\tau\, L \) is invariant under both Poincare $\mathcal{ISO}(1,2)$ spacetime symmetry  and the differomorphism symmetry of $\mathcal{P}_{m_p}$ \cite{AmelinoCamelia:2003xp}. Furthermore, $f(p^2)$ should be determined by the geodesic distance $D:=\sup_{\gamma} \int_0^P {\sqrt{ -g_{\mu\nu}(p') \, dp'^{\mu} dp'^{\nu} }}_{\gamma}$ \footnote{Note that, this manifold is a Lorentzian one (pseudo-Riemannian). We have to, therefore, make use of a supremum, rather than an infimum, to compute the geodesic distance $D$, as this should correspond to the longest possible proper distance between a pair of points separated by timelike intervals. Here, the triangle inequality gets reversed. This is in contrast to the use of infimum in the case of Riemannian manifolds of Euclidean signature. In the case of $AdS_3$ manifold,however, we need to consider, small distances $D$( in comparison to the radius of $AdS_3$) only. This is to avoid any conjugate points, where the geodesics may refocus and thereby jeopardizing the existence of any unique supremum. Further, this restriction, on small momentum is required to avoid the closed timelike curves in the standard $AdS_3$ and not their universal cover.} where $\gamma$ being the timelike trajectory connecting  the origin $P^{\mu}=0$ and an arbitrary point $P\in \mathcal{P}_{m_p}$ with coordinate $p^{\mu}$.}   
Now $D$ can be determined,without making use of the geodesic equation explicitly, by making use of the following differential equations, 
\begin{eqnarray}\label{34}
(\partial_{\mu}C)g^{\mu\nu}(p)(\partial_{\nu}C)=-4C
\end{eqnarray}    
fulfilled by $C=D^2$ which can be regarded as is the modified d'-Alembertian operator. This identity can be proved trivially by using the fact that the both left and right hand sides of this equation are scalars under diffeomorphism of $\mathcal{P}_{m_p}$ and therefore its equality can be verified in any frame of our choice. And for that we can choose the Riemann normal coordinates $\pi^{a}$ to coordinatize $P \in \mathcal{P}_{m_p}$ using the flat tangent space $T_0(\mathcal{P}_{mp})
$ at the origin $(P^{\mu}=0)$ and write 
\begin{eqnarray}\label{35}
C=D^2=-{\eta}_{ab}\pi^{a}\pi^{b} 
\end{eqnarray}
where $\pi^a$ can be written as 
\begin{eqnarray}\label{36}
\pi^a=n^a\sqrt{-f(p^2)}
\end{eqnarray}
with $n^a=(cosh\phi,sinh\phi cos\theta, sinh\phi sin\theta)$ be the unit time like vector $(n^an_a=-1)$ which is tangent to the geodesic at the origin. The above identity can now be verified trivially in this frame. 
Now setting, 
\begin{eqnarray}\label{37}
 C=-f(p^2)
\end{eqnarray} 

by noting  $f(p^2)<0$ we can demand, that it should have the correct commutative limit, i.e., $f(p^2)\rightarrow p^2=-m^2$ when the limit $m_p\rightarrow \infty$ is taken.    
A straight forword computation then yields,
\begin{eqnarray}\label{38}
  f(p^2)=-m_{p}^2\left[ \tan^{-1}\left( \frac{\sqrt{-p^2}}{m_p} \right) \right]^2:=-M^2
\end{eqnarray} 

Here, \( M = m_p \left[ \tan^{-1} \left( \frac{m}{m_p} \right) \right] \) defines the "renormalized" mass in the spirit of ~\cite{Mezincescu2000}, which reduces to \( M \to m \) in the commutative (flat momentum space) limit. This identification reflects the standard notion of mass renormalization in quantum field theory,albeit a finite one in this context. Notably, while the bare mass \( m \) is unbounded from above, the renormalized mass \( M \) is bounded as \( M < (\pi/2)\, m_p \). This implies that the renormalized mass cannot exceed the Planck scale, \( m \lesssim m_p \), for the renormalized description to remain valid.

\section{Effective Energy-Momentum Tensor}

To study the gravitational backreaction sourced by a relativistic
point particle whose momentum space is curved, we now derive the
corresponding effective energy--momentum tensor. This requires
coupling the particle dynamics to a general background metric
$g_{\mu\nu}(q)$, where $q^\mu$ denotes the effective configuration--
space coordinates introduced earlier.

Before proceeding, it is useful to briefly recall the geometric
origin of these coordinates. Our underlying spacetime model is
described by noncommutative generators $x^a$ obeying the Lie--Poisson
algebra
\begin{equation}
\{x^a, x^b\} = \frac{1}{m_p}\epsilon^{abc}x_c .
\end{equation}
This algebra arises as the classical limit of a fuzzy
$\mathbb{R}^{1,2}_\star$ noncommutative geometry and defines a
Poisson manifold whose Poisson tensor is degenerate. As a
consequence, the variables $x^a$ label symplectic leaves
(coadjoint orbits) rather than forming a commuting coordinate
chart on a smooth manifold. Although one may consistently define
Riemannian or Lorentzian structures on individual symplectic
leaves, the noncommuting variables $x^a$ are not suitable for
defining local geometric observables such as energy--momentum
tensors.

For this reason we employ the Bopp--shifted (Darboux) variables
introduced earlier (\ref{142}), which satisfy the canonical Poisson brackets \eqref{141}. These
coordinates therefore furnish a commuting chart on phase space
that can be used to describe particle trajectories and local
spacetime observables.

{As already discussed in the previous section, the use of the variables $q^\mu$ does not eliminate the underlying noncommutative structure. The noncommutative generators $x^a$ are associated with the geometry of curved momentum space and form a nonholonomic basis on the corresponding momentum space. These generators are related to the holonomic configuration--space coordinates $q^\mu$ through the momentum--dependent triad
\begin{equation}
    x^a(p) = q^\mu E_\mu{}^{\ a}(p),
\end{equation}}

{where $E_\mu{}^{\ a}(p)$ defines a frame on momentum space. Similar realizations of Lie-algebra-type noncommutative coordinates—particularly in the case of $\kappa$-Minkowski $(3+1)$-dimensional quantum spacetime—in terms of canonical phase-space variables have been extensively studied from an algebraic perspective in the literature (see, e.g., \cite{Juric:2015aza,Harikumar:2011um,Meljanac:2010et,Juric:2017bpr,qm1b-snfj}). In contrast, the present work focuses on a $(2+1)$-dimensional setting, where we adopt a geometric interpretation in which the underlying classical limit of the quantum spacetime structure is understood as defining a curved momentum-space geometry.}

{In this formulation, the variables $q^\mu$ furnish a holonomic coordinate chart suitable for describing particle trajectories, while the generators $x^a$ correspond to a nonholonomic frame whose orientation depends on the particle momentum. Consequently, the effects of the curved momentum--space geometry do not appear in the coordinates $q^\mu$ themselves, which remain ordinary commuting configuration--space variables. Rather, they enter through the momentum--dependent frame transformation relating $x^a$ and $q^\mu$, together with the deformed symplectic structure and the modified dispersion relation $f(p^2)$ introduced earlier.}

{In the present work we focus on the low--momentum sector, $|p| \ll m_p$, where the magnitude of the particle three--momentum is small compared to the curvature scale of momentum space. In this regime, the curvature of momentum space manifests itself through controlled corrections to the particle dynamics, while the variables $q^\mu$ provide an effective semiclassical configuration--space description suitable for coupling to classical gravity.}

{It is worth noting that, even at the semiclassical level, a formulation directly in terms of the nonholonomic frame defined by the noncommutative variables $x^a$ could naturally lead to momentum--dependent geometries of Finsler or rainbow type, which lie beyond the standard Riemannian framework. In the present work, however, our aim is to retain a Riemannian spacetime description in which the effects of curved momentum space appear through the effective matter sector, rather than through a momentum--dependent spacetime metric.}

{At the quantum level, in more complete noncommutative gravity frameworks—such as those based on Connes' spectral triple formulation \cite{Chamseddine:1996rw,Schucker:2001aa} or deformation quantization—the metric and distance function would instead be constructed directly from operator structures, without introducing commuting coordinates. Within the present semiclassical treatment, however, the variables $q^\mu$ provide a convenient coordinate chart for defining geometric quantities such as the energy--momentum tensor and the Einstein tensor.}




We now construct the effective configuration-space action (see Appendix C for the derivation) by coupling the particle dynamics to a background metric \( g_{\mu\nu}(q) \):
\begin{equation}
S_{\text{eff}}[q(\tau), g_{\mu\nu}] =
\int d\tau \left[
-\alpha(M, m_p) \sqrt{-g_{\mu\nu}(q) \dot{q}^\mu \dot{q}^\nu}
- \beta(M, m_p) \left(-g_{\mu\nu}(q) \dot{q}^\mu \dot{q}^\nu\right)^{5/2}
\right],
\label{ag}
\end{equation}
where the leading-order curvature corrections are given by 
\begin{equation}
\alpha = M \left(1 + \frac{M^2}{3m_p^2} \right), \quad \beta = \frac{M^3}{3m_p^2}.
\end{equation}

It is important to note that the above configuration-space action is obtained after eliminating the auxiliary variables from the first-order formulation. While the original phase-space action is manifestly reparametrization invariant, the reduced form used here corresponds to a non-affine representation of the worldline dynamics. The underlying symmetry structure remains intact and is explicitly preserved in the first-order formulation.

We can now vary the action \eqref{ag} with respect to the trajectory
$\delta q(\tau)$ in order to determine the particle dynamics.
It is clear from the mass dependence of the ratio $\beta/\alpha$
that the resulting equation of motion will depend on the effective
mass parameter $M$. This feature is expected to persist in similar
models involving curved momentum space in higher dimensions,
such as the $(3+1)$--dimensional framework discussed in
Ref.~\cite{nandi2023}.

{It is important to note that the effective Lagrangian appearing in
\eqref{ag} is not homogeneous of degree one in the velocities,
since it contains higher powers of the invariant quantity
$X=-g_{\mu\nu}\dot q^\mu \dot q^\nu$. Consequently, the action does
not possess reparametrization invariance off shell. As a result,
the Euler--Lagrange equations initially take the form of a
non--affine geodesic equation.}


{However, a consistency analysis of the equations of motion shows that
the quantity $X=-g_{\mu\nu}\dot q^\mu \dot q^\nu$ remains constant
along the worldline. As a result, the non--affine term vanishes on
shell and the equation of motion reduces to
\begin{equation}
D_\tau u^\mu = 0 ,
\end{equation}
where $u^\mu=\dot q^\mu$ and $D_\tau$ denotes the covariant derivative
along the worldline,
\[
D_\tau u^\mu = \frac{d u^\mu}{d\tau}
+ \Gamma^\mu_{\alpha\beta}(q)u^\alpha u^\beta .
\]
This is precisely the affine geodesic equation associated with the
Levi--Civita connection of the background metric, with $\tau$ playing the role of affine parameter.Details of the analysis has been provided in Appendix (E).}

{Therefore, at the level of spacetime trajectories in a fixed
background geometry, the motion remains geodesic and the weak
equivalence principle (in the sense of universality of free fall)
is preserved. The effects of curved momentum space instead appear
through the modified dispersion relation, the relation between
momentum and velocity, and the effective stress--energy tensor
sourcing the gravitational field. In the commutative limit
$m_p\rightarrow\infty$, these corrections vanish and the standard
relativistic point-particle dynamics is recovered.}


Returning to the action \eqref{ag}, we obtain the corresponding
Hilbert energy--momentum tensor by taking the functional derivative
of the action with respect to the background metric,
\begin{equation}
T_{\text{eff}}^{\mu\nu}(q(\tau)\mid q)
=
-\frac{2}{\sqrt{-g}}
\frac{\delta S_{\text{eff}}}{\delta g_{\mu\nu}(q)}
\Big|_{g_{\mu\nu}\to\eta_{\mu\nu}} .
\end{equation}

{As discussed in the previous section, the effective action
\eqref{ag} is not reparametrization invariant since the
Lagrangian is not homogeneous of degree one in the velocities.
Consequently the worldline parameter $\tau$ does not correspond
to a gauge degree of freedom. The functional variation with
respect to the metric is therefore performed directly on the
full action which is invariant under diffeomorphism, without imposing any normalization
condition on the velocity such as $\dot q^2=-1$.
Any convenient parametrization of the trajectory may only be
chosen after the stress tensor has been derived.}

Note that,the quantity $T_{\text{eff}}^{\mu\nu}$ is a composite object
constructed from the worldline field $q(\tau)$ and its derivative
$\dot q(\tau)$, evaluated at a generic spacetime point $q$.

Varying the action yields
\begin{equation}\label{var}
\delta S_{\text{eff}}
=
-\frac{1}{2}
\int d\tau
\left[
\frac{\alpha}{\sqrt{-g_{\mu\nu}(q)\dot q^\mu\dot q^\nu}}
+
5\beta
\left(-g_{\mu\nu}(q)\dot q^\mu\dot q^\nu\right)^{3/2}
\right]
\dot q^\mu \dot q^\nu
\,\delta g_{\mu\nu}(q).
\end{equation}

Substituting this into the definition above, we obtain the
energy--momentum tensor
\begin{equation}
\begin{aligned}
T_{\text{eff}}^{\mu\nu}(q(\tau)\mid q)
&=
\int d\tau
\left[
\frac{\alpha}{\sqrt{-g_{\rho\sigma}(q)\dot q^\rho \dot q^\sigma}}
+
5\beta
\left(-g_{\rho\sigma}(q)\dot q^\rho \dot q^\sigma\right)^{3/2}
\right] \\
&\quad \times
\frac{\dot q^\mu \dot q^\nu}{\sqrt{-g}}
\,\delta^{(3)}(q-q(\tau)).
\end{aligned}
\label{bd}
\end{equation}

In the flat limit $g_{\mu\nu}(q)\to\eta_{\mu\nu}$ this reduces to
\begin{equation}\label{energy}
T_{\text{eff}}^{\mu\nu}(q(\tau)\mid q)
=
\int d\tau
\left[
\frac{\alpha}{\sqrt{-\dot q^2}}
+
5\beta(-\dot q^2)^{3/2}
\right]
\dot q^\mu \dot q^\nu
\,\delta^{(3)}(q-q(\tau)),
\end{equation}
where
\begin{equation}
\dot q^2=\eta_{\mu\nu}\dot q^\mu\dot q^\nu .
\end{equation}

This expression represents the effective energy--momentum tensor of
a relativistic point particle incorporating the leading corrections
arising from the curvature of momentum space, and provides the
appropriate source term for the semiclassical Einstein equations
considered in the next section.\\

In the limit \( m_p \to \infty \), where the momentum space becomes flat \( f(p^2) \to p^2  \), the renormalized mass \( M \) reduces to the standard mass \( m \), and the energy-momentum tensor becomes 
\begin{equation}
T_{\mu\nu}(q(\tau)|q) \to \int d\tau\, m\, u_\mu u_\nu\, \delta^{3}(q - q(\tau)).
\end{equation}
This reproduces the familiar result for a relativistic point particle in flat momentum space.

\section{Geometry from the Effective Action and Curved Momentum Source }\label{5}

We now investigate the classical spacetime response to a point particle
whose dynamics are influenced by a curved momentum space. The total
action consists of the Einstein--Hilbert term with a negative
cosmological constant, $\Lambda_c=-1/\ell_{\text{AdS}_3}^2$, together
with the effective configuration--space action \eqref{ag} describing
the particle,
\begin{equation}\label{action}
S=\frac{1}{16\pi G}\int d^3x\,\sqrt{-g}(R-\Lambda_c)
+S^{\text{matter}}_{\text{eff}}[q(\tau),g_{\mu\nu}] .
\end{equation}

Variation of the total action with respect to the background metric
yields Einstein's field equations
\begin{equation}
G_{\mu\nu}+\Lambda_c g_{\mu\nu}=8\pi G\,T^{\text{eff}}_{\mu\nu},
\end{equation}
where the effective energy--momentum tensor
$T^{\text{eff}}_{\mu\nu}$ is given by \eqref{bd}.

{We now focus on the static rest--frame configuration of the particle,
for which $\dot q^\mu=(1,0,0)$. In the usual relativistic point
particle action the worldline parameter coincides with proper time,
since the action is proportional to the proper length of the
trajectory. In the present case the effective action contains higher
powers of the invariant quantity
$X=-g_{\mu\nu}\dot q^\mu\dot q^\nu$ and is therefore not proportional
to the arc length. Consequently the parameter $\tau$ does not
coincide with proper time.}

{As shown in the previous section, the equations of motion imply that
$X=-g_{\mu\nu}\dot q^\mu\dot q^\nu$ remains constant along the
worldline. In the rest frame we may therefore normalize the velocity
such that $\dot q^2=-1$, corresponding to a timelike trajectory.}

Under this normalization the stress--energy tensor simplifies to
\begin{equation}
T^{\mu\nu}_{\text{eff}}(q)
=\int d\tau\,(\alpha+5\beta)\,u^\mu u^\nu
\frac{\delta^{(3)}(q-q(\tau))}{\sqrt{-g}},
\qquad
u^\mu u_\mu=-1 .
\end{equation}

{We emphasize that the effective stress--energy tensor defined above is covariantly conserved,
\begin{equation}
\nabla_\mu T^{\mu\nu}_{\mathrm{eff}} = 0,
\end{equation}
as a direct consequence of spacetime diffeomorphism invariance of the effective action. This follows from the associated Noether identity and holds provided the particle equations of motion are satisfied.}

{It is important to note that this result does not rely on worldline reparametrization invariance, which is not manifest in the present effective description. Rather, spacetime diffeomorphism invariance ensures consistency with the contracted Bianchi identity $\nabla^\mu G_{\mu\nu}=0$.}

{The source is distributional in nature and should be understood in the sense of distributions, corresponding to a localized point particle in spacetime.}

Taking the trace and integrating over time yields the static spatial
energy density
\begin{equation}
T_{\text{eff}}(q)=-Q\,\delta^{(2)}(\vec q),
\qquad
Q=\alpha+5\beta>0 .
\end{equation}
The quantity $Q$ can therefore be interpreted as the effective mass
parameter associated with the point particle inserted in
$AdS_3$ spacetime. And the transformation $M\rightarrow Q$ can be regarded as another 'renormalization'. 

Substituting this source into the trace of Einstein's equations gives
the Ricci scalar
\begin{equation}\label{scalar}
R(q)=-\frac{6}{\ell^2}+16\pi G Q\,\delta^{(2)}(\vec q).
\end{equation}

The delta--function curvature singularity should be understood in the distributional sense and indicates the presence of a conical defect in the spacetime geometry. In asymptotically
$AdS_3$ spacetimes the physical mass is determined from the
asymptotic behavior of the metric at the boundary rather than by
integrating the scalar curvature. The deficit angle produced by the
localized source will be used to determine  the ADM mass $(M_{ADM})$ of the solution which we shall compute in the next subsection. This is the energy measured by an observer, stationed at the boundary of $AdS_3$. As we shall demonstrate that $M_{ADM}$ will have non-linear relation on $Q$, as we need to take gravitational binding energy in consideration.      

The deformation of momentum space will thus introduce an effective
regularization on $M_{ADM}$ like $M$. In fact, as the renormalized mass
$M=m_p\tan^{-1}(m/m_p)$ saturates at large energies, the ADM mass
too will remain finite. This contrasts with the undeformed point particle in
$(2+1)$--dimensional gravity, where the corresponding mass parameter
can diverge.

We emphasize that the mass $Q$ is not gravitational in origin. Rather,
it arises as an effective parameter characterizing the particle
dynamics induced by the curved momentum space. Nevertheless the
Einstein equations respond to $T_{\mu\nu}$ as a classical source.
Importantly, this stress--energy tensor is defined on the same
spacetime manifold whose geometry is being solved for, consistent
with a semiclassical gravitational description. The resulting
spacetime geometry therefore encodes the effects of the curved
momentum space through the backreaction of the effective matter
source.

\section*{5.1 ADM Mass from the Effective Source}

In asymptotically $AdS_3$ spacetimes the physical mass is defined as
an asymptotic conserved charge associated with time translations,
rather than by a bulk integral of the scalar curvature. In the present
case the localized curvature produced by the point particle determines
the conical defect of the geometry, which in turn fixes the mass
parameter $\mu$ (see below) appearing in the $AdS_3$ metric. The ADM mass is then
obtained from the Brown--York~\cite{BrownYork} quasi-local stress tensor, which is
equivalent to the Brown--Henneaux~\cite{BrownHenneaux} asymptotic charge construction.

To begin with note that (\ref{scalar}) can be split into two parts as

\begin{equation}
R(q) = R_{reg} + R_{sing}(q)
\label{1}
\end{equation}

where $R_{reg} = -\dfrac{6}{\ell^{2}}$ is a constant stemming from the cosmological constant and 
$R_{sing} = 16\pi G Q \, \delta^{(2)}(\tilde{x})$ represents the singular curvature associated with the point particle located at the origin. 
Thus $R_{reg}$ is the sole contributor to the curvature away from the origin $(r>0)$.

\vspace{0.2cm}

\noindent
The standard $AdS_{3}$ line element

\begin{equation}
ds^{2} =
-\left(\frac{r^{2}}{\ell^{2}}-\mu\right) dt^{2}
+ \frac{1}{\left(\frac{r^{2}}{\ell^{2}}-\mu\right)} dr^{2}
+ r^{2} d\phi^{2}, \qquad r>0
\label{169}
\end{equation}

reproduces this regular part $R_{reg} = -\dfrac{6}{\ell^{2}}$ in the region $r>0$, irrespective of the value of the dimensionless mass parameter $\mu$. 
If $Q=0$ the above line element can be smoothly extended to include the origin $r=0$ by setting $\mu=-1$ thereby making it geodesically complete. In that case one obtains an empty $AdS_3$ spacetime.

However, the presence of a non–vanishing $Q$ produces a conical deficit angle $\Delta\phi$, which in turn depends on $\mu$. This relation will allow us to determine $\mu$ in terms of $Q$, provided both parameters lie within the allowed threshold.

To see this explicitly, consider the spatial part of the line element (i.e.\ the $t=\text{const}$ surface of (\ref{169})),

\begin{equation}
ds^{2}\big|_{(r,\phi)} =
\frac{1}{\left(\frac{r^{2}}{\ell^{2}}-\mu\right)} dr^{2}
+ r^{2} d\phi^{2}
:= g^{(2)}_{ij}dx^idx^j; \quad i,j \in [1,2].
\label{3}
\end{equation}

Now in the vicinity of $r\approx 0$, one can write for $\mu < 0$

\[
\rho = \frac{r}{\sqrt{-\mu}}
\]

so that

\begin{equation}
ds^{2}\big|_{(r,\phi)} \approx d\rho^{2} + \rho^{2}(-\mu)d\phi^{2}.
\label{4}
\end{equation}

With this parametrization the total angle of rotation becomes $\delta\phi = 2\pi\sqrt{-\mu}$ as $0 \le \phi < 2\pi$. 
Correspondingly the deficit angle is given by

\begin{equation}
\Delta\phi = 2\pi - \delta\phi = 2\pi(1-\sqrt{-\mu}).
\label{5}
\end{equation}

An alternative derivation has been provided in the Appendix (F)  using holonomy arguments. 
Note here that a deficit angle arises as long as $-1 < \mu < 0$. The case $\mu=-1$ corresponds to pure $AdS_{3}$ spacetime, often referred to as the $AdS_{3}$ vacuum in the CFT language. 
For $\mu>0$ there is no deficit angle; instead the geometry corresponds to the BTZ black hole regime.

\vspace{0.2cm}

\noindent
We now relate the effective mass $Q$ with the deficit angle $\Delta\phi$ using the Gauss–Bonnet theorem. 
Integrating over an infinitesimal circular disc $D$ around $r=0$ gives

\begin{equation}
\int_D R_{sing}\sqrt{g^{(2)}}\, d^2x = 2\Delta\phi.
\label{6}
\end{equation}

Using (\ref{1}) this yields

\begin{equation}
\Delta\phi = 8\pi G Q.
\label{7}
\end{equation}

Here we have used the $2D$ diffeomorphism–invariant form of the singular curvature

\begin{equation}
R_{sing}=16\pi G Q\,\dfrac{\delta^{(2)}(\tilde{v})}{\sqrt{g^{(2)}(x)}}.
\label{8}
\end{equation}

Combining (\ref{5}) and (\ref{7}) gives the relation between the effective mass $Q$ and the parameter $\mu$,

\begin{equation}
Q=\dfrac{1}{4G}\left(1-\sqrt{-\mu}\right).
\label{22}
\end{equation}

Equivalently

\begin{equation}
\mu=-(1-4GQ)^2<0.
\label{23}
\end{equation}

Writing this relation in the symmetric form

\begin{equation}
\sqrt{|\mu|}=|1-4GQ|
\end{equation}

makes its analytic structure manifest. This expression smoothly extends across the transition point $Q=1/(4G)$ and naturally produces two branches of solutions corresponding to particle and black hole geometries. In this sense, the parameter $\mu$ determined in the particle regime analytically continues into the BTZ regime, providing a unified description of both conical defect and black hole spacetimes.

The resulting relation can therefore be written compactly as

\begin{equation}
\mu(Q) =
\begin{cases}
-(1-4GQ)^2, & Q < \dfrac{1}{4G}, \\
0, & Q = \dfrac{1}{4G}, \\
(4GQ-1)^2, & Q > \dfrac{1}{4G}.
\end{cases}
\end{equation}

The first branch ($Q<1/(4G)$) corresponds to conical defect geometries,
while the second branch ($Q>1/(4G)$) corresponds to BTZ black holes.
At the critical value

\begin{equation}
Q=\bar{Q} = \frac{1}{4G},
\end{equation}

the deficit angle reaches $\Delta\phi=2\pi$, the conical geometry
pinches off, and a horizon forms. This corresponds to zero mass black hole Beyond this point the spacetime
belongs to the massive BTZ black hole branch with $\mu>0$.

In this context, we would like to point out that this critical value   $\bar{Q}$ which is required for black hole formation, can be rewritten as,  

\begin{equation}
    \bar{Q}=\frac{1}{4G}=\frac{m_p}{4}\label{181}
\end{equation}

where, we have identified the Planck mass $m_p=\frac{1}{G}$. (Appendix A). The corresponding, renormalized mass $\bar{M}$ satisfies the equation, 

\begin{equation}
\bar{Q} = \bar{M}\left(1 + \frac{2\bar{M}^2}{m_p^2}\right)\label{182}
\end{equation}

Now, using (\ref{181}) and (\ref{182}), we can solve for the dimensionless quantity, $X=\frac{\bar{M}}{m_p}$ through the cubic equation, 

\begin{equation}
8X^3+4X-1=0
\end{equation}

This admits only one real root is given by $X=0.227$. Substituting it back, in eq(1), one gets, the corresponding critical value of bare mass $\bar{m}$, given by,

\begin{equation}
\bar{m}=m_ptan X=0.231m_p >\bar{M}
\end{equation}

This shows, bare mass ($\bar{m}$), undergoes successive "renormalization", and in the process its value gets diminished slightly in $\bar{M}$ but subsequently. gets enhanced in $\bar{Q}$ substantially  to facilitate its entry into the black hole regime as we shall see below. 

\vspace{0.2cm}

\noindent
We now compute the ADM mass associated with the geometry \ref{169}. 
This represents the energy measured by an observer located at the boundary of $AdS_3$. 
For this purpose we employ the Brown–York formulation \cite{BrownYork}, which expresses the quasi–local energy in terms of the extrinsic curvature of the spatial boundary at $r\rightarrow\infty$.

The induced metric on this boundary is

\begin{equation}
\left. ds^2 \right|_{(t,\phi)} = h_{ab} dx^a dx^b
      = -f^2(r)dt^2 + r^2 d\phi^2 ,
\qquad
f^2(r)=\left(\frac{r^2}{\ell^2}-\mu\right).
\label{9}
\end{equation}

The outward pointing unit normal vector is

\[
n^\mu = (0,f(r),0),
\]

which satisfies the normalization condition $n^\mu n_\mu = 1$.

The extrinsic curvature is defined through the Lie derivative as, 

\begin{equation}
K_{ab} = \frac{1}{2}\mathcal{L}_n h_{ab}.
\label{10}
\end{equation}

Since the metric is diagonal this simplifies to

\begin{equation}
K_{ab} = \frac{1}{2} n^r \partial_r h_{ab}
       = \frac{1}{2} f(r) \partial_r h_{ab}.
\label{11}
\end{equation}

Componentwise this yields

\begin{equation}
K_{tt} = -\frac{r}{\ell^2} f'(r),
\qquad
K_{\phi\phi} = \frac{f(r)}{r}+f'(r),
\label{12}
\end{equation}

where

\[
f'(r)=\frac{df(r)}{dr}=\frac{r}{\ell^2 f(r)}.
\]

Following Brown and York, the boundary stress tensor is defined as

\begin{equation}
T_{ab}=\frac{1}{8\pi G}
\left(K_{ab}-Kh_{ab}+\frac{1}{\ell}h_{ab}\right),
\qquad K=h^{ab}K_{ab}.
\label{13}
\end{equation}

The last term acts as a counterterm that cancels divergences and originates from the boundary action

\begin{equation}
S_{ct}=-\frac{1}{8\pi G\ell}\int_{\mathcal{B}} \sqrt{-h}\, d^{2}x,
\label{14}
\end{equation}

where $\mathcal{B}$ denotes the boundary of $AdS_3$.

One can verify that this stress tensor is covariantly conserved,

\begin{equation}
\nabla^{a}T_{ab}=0,
\label{15}
\end{equation}

since the metric components $h_{ab}=\text{diag}(-f^2(r),r^2)$ are independent of both $t$ and $\phi$. This is required to be satisfied, in order to ensure that there is no leakage at the boundary. 

Furthermore the trace

\begin{equation}
T=T^{a}_{\ a}=h^{ab}T_{ab}\rightarrow 0
\quad \text{as} \quad r\rightarrow\infty
\label{16}
\end{equation}

ensures that conformal invariance is restored at the boundary.

\vspace{0.2cm}

\noindent
The ADM mass is defined as the conserved charge associated with the Killing vector $\partial_t$,

\begin{equation}
M_{ADM}=\oint \sqrt \sigma\, d\phi \, (u^{a}(\partial_{t})^{b}T_{ab}),
\label{17}
\end{equation}

where $u^{a}=(\frac{1}{f},0)$ is the unit time–like vector satisfying $u^{a}u^{b}h_{ab}=-1$, and $\sqrt{\sigma}d\phi=r\,d\phi$ is the invariant "volume element" of the boundary circle $S^1$.

Taking the limit $r\rightarrow\infty$ yields

\begin{equation}
M_{ADM}=\frac{\mu}{8G}.
\label{18}
\end{equation}

The same result can also be obtained using the Brown–Henneaux asymptotic symmetry analysis \cite{BrownHenneaux}. 
Holographic renormalization studies \cite{BalasubramanianKraus,Skenderis2002} have shown that these two approaches are equivalent.

\vspace{0.2cm}

\noindent
Let us now discuss the physical interpretation of different regimes of $\mu$.

For $\mu=-1$ the metric near the origin becomes

\begin{equation}
ds^{2}\approx -dt^{2}+dr^{2}+r^{2}d\phi^{2},
\label{19}
\end{equation}

demonstrating that,  to pure $AdS_3$ spacetime with no matter present has the structure of flat Minkowski spacetime near $r=0$. 
The associated ground state energy is therefore

\begin{equation}
E_{0}=-\frac{1}{8G} < 0.
\label{20}
\end{equation}

Insertion of an effective mass $Q$ at the origin raises the energy of the spacetime. 
However, the resulting energy is not simply

\begin{equation}
M_{ADM}=Q-\dfrac{1}{8G},
\label{21}
\end{equation}

since one must also take into account the gravitational binding energy \cite{DeserJackiw}.

Using the relation obtained earlier between $Q$ and $\mu$,
\begin{equation}
Q=\dfrac{1}{4G}\left(1-\sqrt{-\mu}\right),
\end{equation}
the excitation energy above the $AdS_3$ vacuum becomes

\begin{equation}
M_{\text{excitation}}
=
\dfrac{\mu}{8G}-\left(-\dfrac{1}{8G}\right).
\label{26}
\end{equation}

Upon simplification this yields

\begin{equation}
M_{\text{excitation}}=Q-2GQ^2,
\label{27}
\end{equation}

where the second term $(-2GQ^2)$ represents the gravitational binding energy.\cite{DeserJackiw}

Finally, when $Q>\frac{1}{4G}$ the geometry enters the BTZ regime. 
The corresponding mass parameter becomes

\begin{equation}
\mu_{BTZ}=(4GQ-1)^2,
\label{28}
\end{equation}

and the associated ADM mass is

\begin{equation}
M_{ADM}=\dfrac{\mu_{BTZ}}{8G}
       =\dfrac{1}{8G}(4GQ-1)^2>0.
\label{29}
\end{equation}

In this regime the singularity at $r=0$ is shielded by an event horizon located at

\begin{equation}
r_h=\ell\sqrt{\mu_{BTZ}}.
\label{30}
\end{equation}

In $(2+1)$-dimensional gravity, there are no local propagating gravitational degrees of freedom. As a result, spacetime is locally flat (or of constant curvature in the presence of a cosmological constant), and all physical information is encoded in global properties. In particular, a point source induces a conical defect whose geometry is completely characterized by its holonomy (see Appendix D for an explicit construction).

The deficit angle $\Delta\phi$ associated with the conical geometry is determined by the strength of the effective source, here proportional to $(\alpha + 5\beta)$. This deficit angle is related to the mass parameter $\mu$ via
Thus, the combination $Q=(\alpha + 5\beta)$ fixes the holonomy, which in turn determines the deficit angle and hence uniquely specifies the ADM mass of the spacetime.

\section*{5.3 Deformed BTZ Black Hole and Thermodynamics}

In $(2+1)$-dimensional gravity with a negative cosmological constant, a point source gives rise to a spacetime that is locally AdS$_3$, with global properties determined by its holonomy. The corresponding metric can be written in BTZ form,
\begin{equation}\label{metric}
ds^2 = -f^2(r)\, dt^2 + \frac{dr^2}{f^2(r)} + r^2\, d\phi^2.
\end{equation}

In the present case, the geometry is determined by a nonlinear relation between the microscopic parameter $M$ and the ADM mass. Defining
\begin{equation}
Q(M) =\alpha+5\beta= M\left(1 + \frac{2M^2}{m_p^2} \right),
\label{v}
\end{equation}
and
\begin{equation}
A_{m_{p}}(M) = 4G\,Q(M) - 1,
\label{w}
\end{equation}
the ADM mass is given by
\begin{equation}
M_{\text{ADM}} = \frac{A_{m_{p}}^2(M)}{8G}.
\label{mn}
\end{equation}

It is useful to relate this parametrization to the standard BTZ form. In the usual description, the geometry is characterized by a parameter $\mu$, which can be written as $\mu = (4GQ - 1)^2$. In the present framework, this quantity can equivalently be expressed as
\begin{equation}
\mu = A^2,
\end{equation}
so that the function $A_{m_p}(M)$ provides a natural generalization of the BTZ parameter in terms of the deformed mass variable.

Substituting this into the BTZ lapse function yields
\begin{equation}
f^2(r) = -8G M_{\text{ADM}} + \frac{r^2}{\ell^2}
= -A_{m_{p}}^2(M) + \frac{r^2}{\ell^2}.
\end{equation}

In the BTZ black hole regime, defined by $4GM>1$, one has $A_{m_p}(M)>0$, and the horizon radius is given by
\begin{equation}
r_+ = \ell\,A_{m_p}(M).
\end{equation}


The surface gravity is
\begin{equation}
\kappa = \frac{1}{2} \left. \frac{d f^2(r)}{dr} \right|_{r = r_+}
= \frac{r_+}{\ell^2},
\end{equation}
and the Hawking temperature becomes
\begin{equation}
T_H = \frac{\kappa}{2\pi} = \frac{r_+}{2\pi \ell^2}.
\end{equation}

Assuming the validity of the Bekenstein--Hawking entropy formula in $(2+1)$ dimensions~\cite{carlip1995b}, the entropy(in the unit $\hbar=k_B=1$) is
\begin{equation}
S = \frac{2\pi r_+}{4G}
= \frac{\pi \ell}{2G}\,A_{m_{p}}(M).
\end{equation}

\vspace{0.2cm}
\noindent

{We emphasize that the appropriate thermodynamic energy is the ADM mass (\ref{mn}), which is determined by the global geometry—equivalently, by the holonomy class (see Appendix~F for instance)—and represents the total energy measured by an observer at the boundary. Although the effective description involves a nonlinear relation between the microscopic parameter $M$ and the geometry, thermodynamic quantities depend solely on $M_{\mathrm{ADM}}$, which captures the gravitational backreaction.}

{Using the standard BTZ relations
\begin{equation}
T_H = \frac{r_+}{2\pi \ell^2}, 
\qquad
S = \frac{2\pi r_+}{4G},
\end{equation}
together with
\begin{equation}
r_+ = \ell \sqrt{8G\,M_{\mathrm{ADM}}},
\end{equation}
we can explicitly verify the first law of black hole thermodynamics. Differentiating, we obtain
\begin{equation}
dS = \frac{2\pi}{4G}\,dr_+,
\qquad
dM_{\mathrm{ADM}} = \frac{r_+}{4G\ell^2}\,dr_+,
\end{equation}
which immediately yields
\begin{equation}
dM_{\mathrm{ADM}} = T_H\, dS.
\end{equation}}

Thus, all Planck-scale corrections are encoded entirely in the nonlinear mapping between the microscopic parameter $M$ and the ADM mass $M_{\mathrm{ADM}}$. The thermodynamic structure itself remains unchanged, retaining the standard BTZ form.

\vspace{0.4cm}

\noindent A comparison with the classical BTZ solution is summarized below:
\begin{center}
\renewcommand{\arraystretch}{1.3}
\begin{tabular}{@{}lll@{}}
\toprule
{Quantity} & {Classical BTZ ($m$)} & {Deformed BTZ ($M$)} \\
\midrule

$A$ 
& $A_{\infty}(m) = 4Gm - 1$ 
& $A_{m_{p}}(M) = 4GM\!\left(1+\dfrac{2M^{2}}{m^{2}_{p}}\right)-1$ \\

ADM mass 
& $\dfrac{A^{2}_{\infty}(m)}{8G}$ 
& $\dfrac{A_{m_{p}}^{2}(M)}{8G}$ \\

Lapse function $f^2(r)$ 
& $\dfrac{r^2}{\ell^2} - A^{2}_{\infty}(m)$ 
& $\dfrac{r^2}{\ell^2} - A_{m_{p}}^{2}(M)$ \\

Horizon radius $r_+$ 
& $\ell\,A_{\infty}(m)$ 
& $\ell\,A_{m_{p}}(M)$ \\

Hawking temperature $T_H$ 
& $\dfrac{A_{\infty}(m)}{2\pi \ell}$ 
& $\dfrac{A_{m_{p}}(M)}{2\pi \ell}$ \\

Entropy $S$ 
& $\dfrac{\pi \ell}{2G}\,A_{\infty}(m)$ 
& $\dfrac{\pi \ell}{2G}\,A_{m_{p}}(M)$ \\

\bottomrule
\end{tabular}
\end{center}



We now analyze the quantum emission process of the backreacted BTZ black hole presented in this work, using the semiclassical tunneling method. Our goal is to determine how the curved momentum space geometry modifies the black hole's radiation spectrum, and whether these corrections impact its lifetime or end state.

\subsection{Hamilton–Jacobi Tunneling Method: A Brief Derivation}

Hawking radiation can be interpreted as a quantum tunneling process in which particles escape across the event horizon. The semiclassical framework developed by Parikh and Wilczek~\cite{Parikh:1999mf} employs the WKB approximation to calculate the imaginary part of the classical action for a particle crossing the horizon. A more detailed analysis of Hawking radiation via tunneling methods, including beyond-semiclassical effects and related aspects, has been carried out by Banerjee et al.~\cite{Banerjee:2008ry, Banerjee:2008cf, Banerjee:2008sn}. For completeness, we briefly summarize the key steps of this approach. In fact This is particularly interesting because quantum tunneling processes and Hawking radiation are known to be affected by noncommutative effects in various settings (see, e.g., \cite{Nicolini:2008,Smailagic:2010,Thom:2009zz}).

Consider a spherically symmetric black hole geometry with metric:
\begin{equation}\label{72}
ds^2 = -f^2(r) dt^2 + \frac{dr^2}{f^2(r)} + r^2 d\phi^2.
\end{equation}
Near the horizon $r = r_+$, the function $f^2(r)$ has a simple zero: $f^2(r) \approx \kappa (r - r_+)$, where $\kappa$ is the surface gravity.

To describe the motion of a massless particle, we use the Hamilton--Jacobi equation
\begin{equation}
g^{\mu\nu} \partial_\mu I \, \partial_\nu I = 0,
\end{equation}
where \(I\) is the classical action. For radial motion, we adopt the ansatz
\begin{equation}
I = -\omega t + W(r),
\end{equation}
where \(\omega\) is the particle’s energy. Substituting into the Hamilton--Jacobi equation yields
\begin{equation}
-f^{-2}(r)\,\omega^2 + f^2(r)\,(W')^2 = 0
\quad \Rightarrow \quad
W'(r) = \frac{\omega}{f^2(r)}.
\end{equation}

The radial part of the action is therefore
\begin{equation}
W(r) = \int \frac{\omega}{f^2(r)}\, dr.
\end{equation}

Near the horizon \(r = r_+\), the metric function behaves as
\begin{equation}
f^2(r) \approx \kappa (r - r_+),
\end{equation}
where \(\kappa\) is the surface gravity. The integral then becomes
\begin{equation}
W(r) \approx \frac{\omega}{\kappa} \int \frac{dr}{r - r_+}.
\end{equation}

The integrand has a simple pole at \(r = r_+\), and thus the integral must be defined by deforming the contour in the complex \(r\)-plane. Implementing the Feynman prescription \(r \to r_+ - i\epsilon\), one obtains
\begin{equation}
\int \frac{dr}{r - r_+ - i\epsilon} = \text{P.V.} + i\pi,
\end{equation}
where P.V. stands for the Principal Value. 
so that the action acquires an imaginary contribution from the pole:
\begin{equation}
\text{Im}\, W = \frac{\pi \omega}{\kappa}.
\end{equation}

{Alternatively, this result can be obtained by writing
\begin{equation}
W(r) \sim \frac{\omega}{\kappa} \ln(r - r_+),
\end{equation}
and noting that crossing the horizon requires an analytic continuation across the branch cut of the logarithm,
\begin{equation}
\ln(r - r_+) \rightarrow \ln|r - r_+| + i\pi.
\end{equation}
Since the temporal part of the action remains real, the imaginary contribution arises entirely from the radial part. One therefore finds
\begin{equation}
\mathrm{Im}\,W = \frac{\pi \omega}{\kappa}, \qquad \text{and hence} \qquad \mathrm{Im}\,I = \mathrm{Im}\,W.
\end{equation}}

The emission probability is therefore
\begin{equation}
\Gamma \sim e^{-2\, \text{Im} I}
= \exp\left(-\frac{2\pi \omega}{\kappa} \right)
= \exp\left(-\frac{\omega}{T_H} \right),
\end{equation}
where the Hawking temperature is
\begin{equation}
T_H = \frac{\kappa}{2\pi}.
\end{equation}

This reproduces the standard Hawking result. However, the crucial refinement of Parikh and Wilczek was to enforce energy conservation: when a particle of energy \(\omega\) is emitted, the black hole mass decreases from \(M\) to \(M - \omega\). Consequently, the background geometry --- and hence the horizon location --- evolves during the emission process. One must therefore compute
\begin{equation}
\text{Im} \, I
= \int_{r_{\text{in}}}^{r_{\text{out}}}
\int_0^\omega
\frac{d\omega'}{f^2(r; M - \omega')} \, dr.
\end{equation}

This expression incorporates the backreaction of the emitted particle on the spacetime geometry. Remarkably, the result can be written in terms of the change in black hole entropy:
\begin{equation}
\Gamma(\omega) \sim \exp\left[\Delta S\right],
\qquad
\Delta S = S(M - \omega) - S(M).
\end{equation}

This shows that the tunneling probability is governed by the difference in the number of accessible microstates before and after emission. The resulting spectrum is therefore not exactly thermal, allowing for correlations between successive emissions and suggesting a possible mechanism for information recovery.

\subsection{Application to the Deformed BTZ Black Hole}

We now apply the tunneling formalism to the deformed BTZ black hole. 
As discussed in Section~\ref{5}, the ADM mass sourcing the geometry is no longer directly identified with the particle mass parameter $M$. Instead, it is determined through a nonlinear relation arising from the curved momentum space structure, given by Eq.~(\ref{v}):

\begin{equation}
M_{\text{ADM}}(M)
=
\frac{1}{8G}
\left[
4GM\!\left(1+\frac{2M^2}{m_p^2}\right)-1
\right]^2.
\end{equation}

The corresponding Bekenstein--Hawking entropy is
\begin{equation}
S(M_{\text{ADM}}) = \frac{\pi \ell}{2G}\sqrt{8GM_{\text{ADM}}}.
\end{equation}

In the tunneling picture, energy conservation requires that under the emission of a quantum of 
energy $\omega$, the ADM mass decreases as
\begin{equation}
M_{\text{ADM}} \;\longrightarrow\; M_{\text{ADM}} - \omega.
\end{equation}
Equivalently, if one prefers to parameterize the state in terms of the underlying parameter $M$, 
the new value $M'$ is defined implicitly by
\begin{equation}
M_{\text{ADM}}(M') = M_{\text{ADM}}(M) - \omega.
\end{equation}
Expanding to leading order in $\omega$, we obtain
\begin{equation}
M' \;\approx\; 
M 
-
\frac{\omega}{
\Bigl[4GM\!\left(1+\frac{2M^2}{m_p^2}\right)-1\Bigr]
\left(1+\frac{6M^2}{m_p^2}\right)
}.
\end{equation}

The resulting entropy change is
\begin{equation}
\Delta S \;=\; S(M_{\text{ADM}}-\omega) - S(M_{\text{ADM}})
\simeq -\omega\,\frac{dS}{dM_{\text{ADM}}},
\end{equation}
with
\begin{equation}
\frac{dS}{dM_{\text{ADM}}} = \frac{2\pi \ell}{\sqrt{8GM_{\text{ADM}}}}.
\end{equation}

Thus the corrected emission rate is
\begin{equation}
\Gamma(\omega) \;\sim\; \exp\!\left[- \omega \cdot \frac{2\pi \ell}{\sqrt{8GM_{\text{ADM}}}} \right],
\end{equation}
which corresponds to the Hawking temperature written in terms of the parameter $M$:
\begin{equation}
T_H(M) 
=
\frac{1}{2\pi \ell}
\left[
4GM\!\left(1+\frac{2M^2}{m_p^2}\right)-1
\right].
\end{equation}

which makes explicit the sequence of Planck-scale corrections to the BTZ temperature.

All such corrections are therefore encoded in the nonlinear map 
\begin{equation}
M \;\longrightarrow\; Q(M) \;\longrightarrow\; M_{\text{ADM}}(Q),
\end{equation}
ensuring that energy conservation is implemented consistently in the tunneling framework.

\subsection{Return time of a probe between the horizon and the $AdS_3$ boundary in curved momentum space}

To understand the impact of a BTZ black hole on matter and radiation in its surrounding spacetime, we consider the propagation of a massless probe in this curved background. Such a probe may be regarded as a photon or any massless test particle. We emphasize that its own gravitational field is neglected, so that it propagates purely as a probe of the geometry.

A particularly relevant observable is the \emph{return time} (or \emph{bouncing time}), defined as the time required for an outgoing light ray or perturbation to travel from a point near the black hole horizon to the asymptotic $AdS_3$ boundary and back. This time scale characterizes how rapidly radiation emitted near the horizon returns after reflection at the boundary, thereby providing insight into equilibration in the black hole--AdS system.

{We emphasize that the time coordinate $t$ appearing in the metric corresponds to the asymptotic AdS$_3$ time. Under Brown--Henneaux boundary conditions, the Killing vector $\partial_t$ is fixed and generates time translations at the conformal boundary. Consequently, $t$ admits a direct physical interpretation as the time measured by an observer at infinity, and the ADM mass is defined as the corresponding conserved charge.\footnote{In asymptotically AdS$_3$ spacetimes, the Brown--Henneaux \cite{BrownHenneaux} boundary conditions fix the normalization of the time-translation generator, ensuring that $t$ defines a gauge-invariant observable associated with boundary time evolution.}}





We begin by considering radial null geodesics. Setting $d\phi = 0$ and $ds^2 = 0$ in the metric
\begin{equation}
ds^2 = - f^2(r)\,dt^2 + \frac{dr^2}{f^2(r)},
\end{equation}
we obtain
\begin{equation}
\frac{dr}{dt} = \pm f^2(r),\label{1460}
\end{equation}
where
\begin{equation}
f^2(r) = \frac{r^2}{\ell^2} - 8G M_{\rm ADM}(M).
\end{equation}

In the presence of curved momentum space, the ADM mass acquires a deformation and can be written as
\begin{equation}
M_{\rm ADM}(M) = \frac{1}{8G} \left[ 4GM\!\left(1+\frac{2M^2}{m_p^2}\right)-1 \right]^2.
\end{equation}
For convenience, we introduce
\begin{equation}
A_{m_{p}}(M) = 4GM\!\left(1+\frac{2M^2}{m_p^2}\right)-1.
\end{equation}

The condition for the existence of a horizon is determined by the ADM mass through $4GM>1$. In the present framework, the effective mass $M$ is related to the source parameter $m$ by
\begin{equation}
M = m_p \tan^{-1}\!\left(\frac{m}{m_p}\right),
\end{equation}
which, in the perturbative regime $m^2/m_p^2 \ll 1$, reduces to
\begin{equation}
M = m - \frac{1}{3}\frac{m^3}{m_p^2} + \mathcal{O}\!\left(\frac{m^5}{m_p^4}\right).
\end{equation}
Since $M<m$ due to Planck-scale corrections, the condition $4GM>1$ implies $4Gm>1$, while the converse does not necessarily hold. This reflects a shift in the critical threshold for black hole formation induced by the underlying momentum-space deformation.

In the BTZ black hole regime, defined by $4GM>1$, one has $A_{m_p}(M)>0$. The horizon radius then follows from $f^2(r_+)=0$ as
\begin{equation}
r_+^2 = \ell^2 A_{m_p}^2(M),
\end{equation}
which gives
\begin{equation}
r_+ = \ell\,A_{m_p}(M).
\label{dd}
\end{equation}

The one-way travel time from a radial position $r_0$ (close to the horizon) to the boundary is obtained by integrating along the null trajectory:
\begin{equation}
\tau_{\rm out} = \ell^2 \int_{r_0}^{\infty} \frac{dr}{r^2 - r_+^2}.
\end{equation}
Evaluating the integral gives
\begin{equation}
\tau_{\rm out} = \frac{\ell^2}{2r_+} \left[ \ln\left(\frac{r - r_+}{r + r_+}\right) \right]_{r_0}^{\infty}.
\end{equation}
Since the contribution at infinity vanishes, the total return time (outgoing plus incoming) becomes
\begin{equation}
\tau_{\rm ret}(M) = \frac{\ell^2}{r_+} \ln\!\left( \frac{r_0 + r_+}{r_0 - r_+} \right).
\end{equation}

Using (\ref{dd}), this can be written explicitly as
\begin{equation}
\tau_{\rm ret}(M) = \frac{\ell}{A_{m_{p}}(M)} \ln\!\left( \frac{r_0 + \ell A_{m_{p}}(M)}{r_0 - \ell A_{m_{p}}(M)} \right).
\end{equation}

In the near-horizon regime, it is convenient to parametrize the emission point as $r_0 = r_+(1+s)$ with fixed $s>0$, which simplifies the result to
\begin{equation}
\tau_{\rm ret}(M) = \frac{\ell}{A_{m_{p}}(M)} \ln\!\left(\frac{2+s}{s}\right).
\end{equation}

This logarithmic behavior is characteristic of near-horizon propagation and reflects the universal time-delay structure of black hole geometries.~\cite{Shapiro:1964}

It is instructive to compare this result with the classical (undeformed) case, for which the horizon radius is
\begin{equation}
r_+^{\rm (class)} = \ell (4Gm - 1),
\end{equation}
where we assume $4Gm>1$ for consistency with the black hole regime. The ratio of return times is therefore
\begin{equation}
\frac{\tau_{\rm ret}^{\rm (mod)}}{\tau_{\rm ret}^{\rm (class)}} = \frac{4Gm - 1}{4GM(1+2M^2/m_p^2)-1}.
\label{z}
\end{equation}

Using the perturbative expansion $M = m - \frac{1}{3}\frac{m^3}{m_p^2}$, this becomes
\begin{equation}
\frac{\tau_{\rm ret}^{\rm (mod)}}{\tau_{\rm ret}^{\rm (class)}} 
= 
\frac{4Gm - 1}{4Gm - 1 + \frac{4Gm^3}{m_p^2}}.
\end{equation}

In the regime $m^2/m_p^2 \ll 1$, and away from the critical point $4GM = 1$, we can expand
\begin{equation}
\frac{\tau_{\rm ret}^{\rm (mod)}}{\tau_{\rm ret}^{\rm (class)}} 
\simeq 
1 - \frac{4Gm^3}{m_p^2(4Gm - 1)} 
+ \mathcal{O}\!\left(\frac{m^4}{m_p^4}\right).
\label{w}
\end{equation}

This expression exhibits a pole at $4Gm = 1$, indicating a strong enhancement of the correction in this regime (see Fig.~2). This divergence arises within the perturbative expansion in terms of $m$ and should be understood as an approximate signal of the underlying geometric transition, which is more precisely determined by the condition $4GM = 1$. Since $M$ differs from $m$ by Planck-scale corrections, the location of the pole expressed in terms of $m$ does not exactly coincide with the true critical point, but instead reflects its vicinity.

\begin{figure}[h!]
\centering
\includegraphics[width=0.75\linewidth]{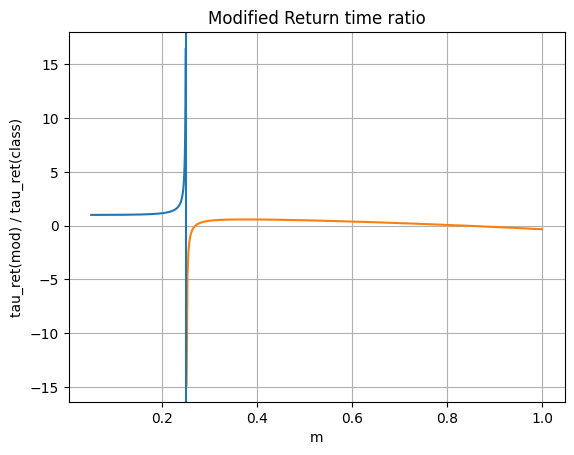}
\caption{Modified return time ratio as a function of mass $m$. The blue branch ($m < \frac{1}{4G}$) corresponds to the particle (conical defect) sector, whereas the orange branch ($m > \frac{1}{4G}$) describes the BTZ black hole regime. The pole at $m = \frac{1}{4G}$ marks the transition between the two sectors. In the plot, we have set $G = 1$ and $m_p = 1$.}
\label{fig:tau_ratio}
\end{figure}

More generally, this result shows that the modification of the return time originates entirely from the deformation of the effective mass entering the geometry. The structure of null geodesics remains unchanged; rather, curved momentum space modifies the relation between the particle parameter $M$ and the spacetime geometry, thereby altering signal propagation times.

Finally, although proper time is not defined for null trajectories, one may still introduce an affine parameter along the geodesic. In the present geometry, this parameter diverges near the horizon (see Appendix~D), reflecting the presence of an infinite redshift surface. In the next subsection, we extend this analysis by incorporating Hawking radiation, allowing the ADM mass to evolve dynamically and thereby further modifying the return time.


\subsection{Return Time of an Emitted Hawking Quantum with Backreaction on the BTZ Geometry}

We now reinterpret the massless particle traveling between the horizon and the $\mathrm{AdS}_3$ boundary as an \emph{emitted Hawking quantum} of energy $\omega$, rather than as a passive probe. In this setting, energy conservation requires that the background geometry responds dynamically through a reduction of the ADM mass,
\begin{equation}
M_{\mathrm{ADM}}(M) \;\longrightarrow\; M_{\mathrm{ADM}}(M) - \omega.
\end{equation}

The ADM mass and horizon radius are given by
\begin{equation}
M_{\mathrm{ADM}}(M)=\frac{A_{m_p}^2(M)}{8G},
\qquad
r_{+}=\ell\, A_{m_p}(M),
\end{equation}
where we assume the BTZ regime $4GM>1$. After the emission of a quantum of energy $\omega$, the horizon radius shifts to
\begin{equation}
r_{+}(\omega)=\ell\sqrt{A_{m_p}^2(M)-8G\omega},
\end{equation}
which captures the backreaction of the Hawking process on the geometry.

{It may be noted that throughout this analysis we work in a controlled perturbative regime characterized by the following assumptions: (i) $\omega \ll M_{\mathrm{ADM}}$, allowing for a perturbative treatment of Hawking backreaction; (ii) $m^2/m_p^2 \ll 1$, ensuring the validity of the expansion of the effective mass $M$ in terms of the source parameter $m$; and (iii) a near-horizon expansion of the emission point $r_0 = r_+(\omega)(1+s)$ with fixed $s>0$ and $s \ll 1$. Higher-order terms in these small parameters are consistently neglected.}

Proceeding as in the previous subsection, the modified return time for the null trajectory from $r_0>r_+(\omega)$ to the boundary and back is
\begin{equation}
\tau_{\text{ret}}(\omega)
=
2\int_{r_0}^{\infty}
\frac{dr}{\dfrac{r^{2}}{\ell^{2}} - 8G\big(M_{\mathrm{ADM}}(M)-\omega\big)}
=
\frac{\ell^{2}}{r_{+}(\omega)}
\ln\!\left(\frac{r_{0}+r_{+}(\omega)}{r_{0}-r_{+}(\omega)}\right).
\end{equation}

Parameterizing $r_0 = r_+(\omega)(1+s)$ with $s \ll 1$, we obtain the near-horizon expression
\begin{equation}
\tau_{\text{ret}}(\omega)
=
\frac{\ell^{2}}{r_{+}(\omega)}
\ln\!\left(1+\frac{2}{s}\right).
\end{equation}

In the regime $\omega \ll M_{\mathrm{ADM}}$, we expand
\begin{equation}
r_{+}(\omega)
=
\ell A_{m_p}(M)
\left(
1 - \frac{4G\omega}{A_{m_p}^2(M)} + \mathcal{O}(\omega^2)
\right).
\end{equation}

Substituting into the return time,
\begin{equation}
\tau_{\text{ret}}(\omega)
\simeq
\frac{\ell}{A_{m_p}(M)}
\left(
1 + \frac{4G\omega}{A_{m_p}^2(M)}
\right)
\ln\!\left(1+\frac{2}{s}\right),
\end{equation}
or equivalently,
\begin{equation}
\tau_{\text{ret}}(\omega)
\simeq
\tau_{\text{ret}}(M)
\left(
1 + \frac{4G\omega}{A_{m_p}^2(M)}
\right).
\end{equation}

{At this stage, it is useful to clearly separate two distinct physical effects contributing to the return time:}
\begin{itemize}
\item[(i)] \textit{Geometric (background) modification:} arising from curved momentum space, which alters the relation between the source parameter $m$ and the effective mass $M$, thereby modifying the background geometry;

\item[(ii)] \textit{Dynamical backreaction:} arising from Hawking emission, encoded through the energy $\omega$, which dynamically changes the ADM mass and horizon radius.
\end{itemize}

We treat these contributions separately before combining them below.

Independently, curved momentum-space effects (without backreaction) modify the return time as
\begin{equation}
\frac{\tau_{\rm ret}^{\rm (mod)}(M)}{\tau_{\rm ret}^{\rm (class)}(m)}=\epsilon,
\end{equation}
with
\begin{equation}
\epsilon= \frac{4Gm - 1}{4GM\left(1+2M^2/m_p^2\right)-1}.
\end{equation}

The relative correction due to Hawking emission is
\begin{equation}
\frac{\tau_{\text{ret}}^{\text{mod}}(\omega) - \tau_{\text{ret}}^{\text{mod}}(M)}
{\tau_{\text{ret}}^{\text{mod}}(M)}
\simeq
\frac{4G\omega}{A_{m_p}^2(M)}.
\end{equation}

Using $\tau_{\text{ret}}^{\text{mod}}(M)=\epsilon\,\tau_{\rm ret}^{\rm (class)}(m)$,
\begin{equation}
\frac{\tau_{\text{ret}}^{\text{mod}}(\omega) - \epsilon\,\tau_{\rm ret}^{\rm (class)}(m)}
{\epsilon\,\tau_{\rm ret}^{\rm (class)}(m)}
\simeq
\frac{4G\omega}{A_{m_p}^2(M)}.
\end{equation}

Now expressing
\begin{equation}
M = m - \frac{1}{3}\frac{m^3}{m_p^2},
\end{equation}
and defining
\begin{equation}
A_{m_p}(m) := A_{m_p}(M(m)),
\end{equation}
we obtain
\begin{equation}
A_{m_p}(m)
=
4Gm - 1 + \frac{4Gm^3}{m_p^2}
+ \mathcal{O}\!\left(\frac{m^5}{m_p^4}\right),
\end{equation}
and
\begin{equation}
\epsilon
\simeq
1 - \frac{4Gm^3}{m_p^2(4Gm - 1)}
+ \mathcal{O}\!\left(\frac{m^4}{m_p^4}\right).
\end{equation}

Combining both effects, the total fractional correction becomes

\begin{equation}
\frac{\tau_{\text{ret}}^{\text{mod}}(\omega) - \tau_{\text{ret}}^{\text{class}}}
{\tau_{\text{ret}}^{\text{class}}}
\simeq
-\underbrace{\frac{4Gm^3}{m_p^2(4Gm-1)}}_{\text{purely geometric effect (curved momentum space)}}
+
\underbrace{\frac{4G\omega}{(4Gm - 1)}}_{\text{effect of Hawking backreaction}} +\mathcal{O}\!\left(\frac{\omega}{m_p^2}\right).
\end{equation}



The final expression clearly separates the two contributions. The first term represents a purely geometric correction induced by curved momentum space, independent of the emission energy $\omega$, while the second term arises from Hawking backreaction and depends explicitly on $\omega$. In the absence of momentum-space deformation ($m_p \to \infty$), the geometric correction disappears, and the return-time modification is governed entirely by the frequency-dependent backreaction term, which remains finite across the transition point. Once Planck-scale deformation is included, however, the geometric contribution develops a pole at $4Gm = 1$, signaling strong sensitivity near the transition. By contrast, the Hawking-induced correction remains finite in this regime.

We thus observe a clear interplay between two physically distinct effects: curved momentum-space deformation tends to decrease the return time through modification of the background geometry, while Hawking backreaction increases it through a dynamical reduction of the horizon radius. Near the transition point, the deformation-induced contribution dominates the leading behavior of the return-time observable.


\section{Conclusion}

To summarize, we have developed a semiclassical framework in which the curvature of momentum space—emerging from an underlying noncommutative spacetime algebra—induces observable corrections to classical gravitational backgrounds. Starting from a fuzzy $\mathbb{R}^{1,2}_\star$ algebra, we demonstrated that its classical limit yields a Lie--Poisson phase space and an emergent AdS$_3$ momentum space geometry. This structure deforms the particle’s dispersion relation and produces an effective configuration-space action with Planck-suppressed corrections. These corrections modify the effective energy-momentum tensor sourcing gravity, while preserving the form of Einstein’s equations.

Our central result is that this deformed stress tensor backreacts on spacetime in a nontrivial and structured manner. Rather than corresponding to a simple renormalization of parameters, the backreaction induces a nonlinear mapping
\[
M \;\longrightarrow\; Q(M) \;\longrightarrow\; A(M) \;\longrightarrow\; M_{\text{ADM}},
\]
which determines the global spacetime geometry. In $(2+1)$ dimensions, where gravity has no propagating local degrees of freedom, this relation is most naturally characterized through holonomies. As we have shown explicitly (see Appendix~V), the conjugacy class of the holonomy uniquely fixes the geometry, allowing us to identify the resulting spacetime with the BTZ family and extract the corresponding ADM mass in a fully geometric manner. In this sense, the modified BTZ solution arises from a consistent matching between source and geometry, rather than from an \emph{ad hoc} parameter substitution.

While the emergence of AdS$_3$ momentum space in 2+1D gravity has also been explored by Amelino-Camelia \textit{et al.}~\cite{Amelino-Camelia:2024tdk}, where it arises from integrating out gravitational degrees of freedom in the Chern--Simons formulation, our approach differs both methodologically and conceptually. Instead of presupposing a group-valued momentum space from a gauge-theoretic action, we start from a noncommutative spacetime algebra and derive the phase space structure by enforcing closure under Jacobi identities. We identify the resulting momentum space geometry with AdS$_3$, where spacetime coordinates emerge as real-valued Killing vector fields of a well-defined metric on the curved momentum space. This geometric reinterpretation allows us to define geodesic distances, construct an effective particle action, and compute the associated stress-energy tensor, which acts as a source in the Einstein equations. Consequently, this leads directly to gravitational backreaction and thermodynamic corrections, now understood as arising from the nonlinear relation between particle parameters and global spacetime charges.

Proceeding further, we have also analyzed the semiclassical emission spectra of the resulting BTZ black hole. Here too, corrections emerge entirely from the curved momentum space geometry, without the need to invoke any \emph{ad hoc} cutoff functions. The tunneling process is governed by the ADM mass obtained from the geometric construction, ensuring that energy conservation is implemented consistently at the level of the spacetime. The key outcomes are: (i) the effective Hawking temperature receives controlled corrections determined by the nonlinear mapping between $M$ and $M_{\text{ADM}}$, and (ii) the emission spectrum acquires non-thermal features via $\Gamma \sim e^{+\Delta S}$ with $\Delta S<0$, indicating the presence of correlations between emitted quanta. In addition, global observables such as the return time of null signals provide an independent probe of the deformation, revealing a nontrivial interplay between kinematic corrections and gravitational backreaction.

\section{Discussion and Outlook}

{The present framework should be contrasted with existing approaches to BTZ geometries and their semiclassical modifications. On the one hand, overspinning BTZ spacetimes have been extensively studied at the classical level, where their causal structure and geodesic properties are well understood despite the presence of a naked singularity~\cite{Briceno:2021}. On the other hand, semiclassical backreaction based on quantum fields introduces corrections through the expectation value of the stress-energy tensor $\langle T_{\mu\nu} \rangle$, but in certain regimes—such as overspinning geometries—this approach can become ill-defined due to divergences that obstruct a consistent perturbative solution~\cite{Baake:2023}.}

We note that in some analyses of near-horizon extremal BTZ geometries, coordinate deformations are introduced as a technical regularization to render asymptotic charges well-defined~\cite{Ballav:2025xyz}. Such deformations are primarily technical in nature and do not introduce new dynamical modifications of physical observables. They should therefore be clearly distinguished from the deformation considered in the present work, which arises from a nontrivial phase-space structure and leads to genuine modifications of physical quantities.

In contrast, the present work adopts a fundamentally different perspective. Rather than introducing quantum fields on a fixed background or employing coordinate regularizations, we incorporate weak quantum spacetime effects through a deformation of particle kinematics induced by curved momentum space. This deformation, rooted in an underlying noncommutative spacetime algebra, leads to an effective stress-energy tensor that remains finite and well-controlled. As a result, the backreaction in our framework manifests as a nonlinear modification of global geometric quantities—such as the ADM mass—while preserving the structural form of the spacetime geometry.

Although developed within the context of $(2+1)$-dimensional gravity—which is characterized by the absence of local gravitational degrees of freedom—our model provides a robust and analytically tractable platform to examine how Planck-scale kinematic deformations, such as curved momentum space, impact classical geometry and black hole thermodynamics. The topological nature of gravity in this reduced dimensionality allows for an exact characterization of backreaction through global quantities such as holonomies and conserved charges, enabling precise calculations of how modified particle dynamics reshape spacetime.

Key phenomena revealed by our approach—such as the nonlinear relation between particle parameters and ADM mass, and the resulting modifications of global observables—are not merely artifacts of the lower-dimensional setting. Rather, they reflect general semiclassical consequences of nontrivial momentum space geometry, which are expected to qualitatively extend to $(3+1)$ dimensions. However, quantitative features such as Hawking radiation spectra and black hole evaporation will differ due to the presence of propagating gravitational degrees of freedom and more intricate horizon structures in higher dimensions.

Thus, the $(2+1)$D framework functions as an effective conceptual and computational laboratory for probing how noncommutative geometry and curved momentum space can regulate ultraviolet behavior and imprint quantum gravitational signatures onto classical spacetimes.

Our construction aligns with a broader spectrum of quantum gravity-inspired models, including doubly special relativity (DSR), $\kappa$-Poincaré symmetries, and the principle of relative locality~\cite{Amelino-Camelia:2011lvm}, which often assume curved momentum spaces formed via group manifolds or quantum deformations. Distinctively, our method originates from a noncommutative spacetime algebra, deriving the curved momentum geometry through algebraic consistency and Jacobi identity closure. This yields a well-defined AdS$_3$ momentum space metric, enabling direct evaluation of geometric quantities such as distances, geodesics, and curvature, while systematically incorporating corrections to semiclassical observables such as energy-momentum tensors, black hole thermodynamics, and signal propagation.

Looking forward, several plausible extensions emerge naturally from our framework:
\begin{itemize}
    \item \textbf{Multiparticle dynamics and curved momentum addition:} Since our phase space structure deforms single-particle kinematics through a nonlinear momentum space geometry, an important next step is to study the addition of momenta in interacting systems. This may reveal nontrivial composition laws and potential links to the principle of relative locality in a concrete $(2+1)$D setting.

    \item \textbf{Entropy corrections via quantum geometry:} Given that our construction modifies the effective stress tensor and associated thermodynamic quantities, it is natural to investigate whether analogous corrections arise in entanglement entropy across horizons, and how momentum space curvature influences information-theoretic aspects of semiclassical gravity.

    \item \textbf{Coupling to spin and internal symmetry:} Extending the framework to include spinning particles in $(2+1)$D may reveal additional geometric structures, particularly through modifications of holonomies in the Chern--Simons formulation.

    \item \textbf{Comparison with existing $(3+1)$D constructions:} Recent work~\cite{Amelino-Camelia:2024tdk} has explored AdS momentum space in higher dimensions. Our results suggest an alternative route based on noncommutative spacetime algebras, which may provide new insights into momentum space geometry beyond group-manifold constructions.

    \item \textbf{Toward an operator-algebraic description of phase space geometry:} While a full spectral triple construction in $(3+1)$D remains highly nontrivial, the structures identified here—Lie--Poisson algebra, emergent vielbein, and curved momentum space metric—offer a promising semiclassical starting point for exploring operator-algebraic formulations of phase space geometry.
\end{itemize}

Overall, our results highlight how a purely kinematical structure—curved momentum space—can consistently imprint itself on classical gravitational observables through global geometric data such as holonomies and conserved charges. This provides a coherent pathway toward bridging noncommutative geometry, quantum gravity remnants, and semiclassical gravitational dynamics.

\section*{Acknowledgements}

The authors would like to thank the anonymous referees for constructive and insightful comments that significantly improved the quality of the manuscript. PN gratefully acknowledges support from the Rector’s Postdoctoral Fellowship Program (RPFP) at Stellenbosch University. He also thanks Fedele Lizzi for insightful comments and suggestions that improved the manuscript, Francesco Petruccione, Director of NITheCS, for continued academic support, and Prof.\ Salvatore Mignemi for insightful discussions and valuable comments during his visit to the University of Cagliari, Italy.
MR acknowledges useful discussions with Abhijit Bandyopadhyay. LH acknowledges financial support from Stellenbosch University through a doctoral fellowship. The authors also thank Anwesha Chakraborty for her collaboration during the early stages of this work.

\appendix
\section*{Appendices}
\addcontentsline{toc}{section}{Appendices}

\renewcommand{\thesection}{\Alph{section}}

\section{Planck Mass and Planck length in $(1+2)D$} 
In order to identify Planck mass and length scale, in $(2+1)D$ we need to overcome  the hurdle stemming from  the absence of Newtonian limit of Einstein gravity in $(2+1)D$, as there is no propagating degrees of freedom; it is just a topological theory and can be rewritten in terms Non Abelian $ISO(1,2)$ Chern-Simons theory . Nevertheless, it has been shown by BTZ~\cite{Banados:1992wn} that one can write down Einstein-Hilbert action in $(2+1)D$ augmented with a negative cosmological term ($\Lambda_c$) as it occurs in the first term in \eqref{action} which admits black hole solutions along with event horizon just in $(3+1)D$ case (For a review see ~\cite{carlip}). In particular the charge-less and non-rotating solution of Einstein's equation following from~\eqref{action}can be written in the following "Schwarzschild form" as we have already used ~\eqref{169} (henceforth we set $c=1$) 
\begin{eqnarray}
ds^2=-f^2(r)dt^2+f^{-2}(r)dr^2+r^2d\phi^2
\end{eqnarray}
where

\begin{equation}
    f^2(r)=\left(\frac{r^{2}}{\ell^{2}}-\mu\right)
\end{equation}. 

Here, as discussed in Section~5.1, $\mu$ is a dimensionless parameter. In the present analysis, we restrict ourselves to the regime $\mu > 0$, which corresponds to the BTZ sector.

To restore physical units, we reintroduce the dimensionful Planck mass $m_p$ and Planck length $l_p$ appropriate to $(2+1)$-dimensional gravity. With these scales reinstated, the modified expression (lapse function) takes the following form:

\begin{eqnarray}\label{mass}
f(r)=\sqrt{-8GM+\frac{r^2}{l^2}};\\\ \label{length}
 \Lambda_c = \frac{1}{l^2}
\end{eqnarray}

As in $(1+3)D$,one can therefore go ahead to  determine the Schwarzschild radius by setting $f(r)=0$ and equating it with the corresponding Compton wavelength to get
\begin{eqnarray}
 l_p\sqrt{Gm_p}&=&\frac{\hbar}{m_p}
 \end{eqnarray} 
where we have identified the mass $M$ in \eqref{mass} with the Planck mass $m_p$ and the length scale $l$ obtained in \eqref{length} through the reciprocal of the  cosmological constant $\Lambda$ with Planck length $l_p$. Note that we have  ignored the numerical factor $8$. This yields the following expressions for the Planck mass($m_p$) and the Planck length $(l_p)$ 
\begin{eqnarray}
m_p=\frac{1}{G};\quad  
l_p=\hbar G
\end{eqnarray}

\section{Geometric Quantization of the Modified Phase Space}

In this appendix we provide a geometric quantization derivation of the quantum operators appearing in the main text. 
We start from the modified phase-space Poisson algebra
\begin{align}
\{x^a,x^b\} &= \frac{1}{m_p}\epsilon^{ab}{}_{c}x^c, \\
\{p^\mu,p^\nu\} &= 0, \\
\{x^a,p^\mu\} &= (E^{-1}(p))^{a}{}^{\mu},
\end{align}
where $(E^{-1}(p))^{a\mu}$ denotes the momentum-space triad (dreibein) and its inverse is denoted by $(E(p))_{a}{}_{\mu}$. 
Indices $a,b,c,\ldots$ denote Lorentz indices, while $\mu,\nu,\ldots$ denote world indices.

A generic phase-space point is written as
\begin{equation}
\xi^A := (x^a,p^\mu),
\end{equation}
and the Poisson structure is encoded in the Poisson bivector
\begin{equation}
J^{AB}=\{\xi^A,\xi^B\}.
\end{equation}

Explicitly,
\begin{equation}
J^{AB}=
\begin{pmatrix}
\dfrac{1}{m_p}\epsilon^{ab}{}_{c}x^c & (E^{-1}(p))^{a\mu} \\
-(E^{-1}(p))^{b\mu} & 0
\end{pmatrix},
\end{equation}
where each block is $3\times3$, so that $J^{AB}$ is a $6\times6$ matrix. 

At this stage, the algebraic structure is defined on the full phase space with coordinates $(x^a,p^\mu)$. However, phase space is not the manifold on which we aim to introduce a metric geometry. Our goal is instead to extract an effective geometry on \emph{momentum space}, which is a manifold in its own right. Accordingly, we restrict attention to the subalgebra of phase-space functions depending only on the momenta, $F=F(p)$. This restriction is consistent, since the Poisson brackets imply
\begin{equation}
\{x_a,F(p)\} = (E^{-1})_{a}{}^{\mu}(p)\,\partial_\mu F(p).
\end{equation}

Greek indices $\mu,\nu,\ldots\in\{0,1,2\}$ denote coordinates $p^\mu$ on momentum space $P_{m_p}$, while Latin indices $a,b,\ldots$ label components with respect to an orthonormal (generally non-holonomic) frame. The objects $(E^{-1})_{a}{}^{\mu}(p)$ define a triad relating the frame $\{x_a\}$ to the coordinate basis $\{\partial/\partial p^\mu\}$,
\begin{equation}
T(P_{m_p})=\mathrm{Span}\{x_a\}
=\mathrm{Span}\!\left\{\frac{\partial}{\partial p^\mu}\right\}.
\end{equation}

A standard result of Poisson geometry states that the map $f\mapsto X_f$, defined by $X_f(\cdot)=\{f,\cdot\}$, associates to each function a vector field. Moreover, the Jacobi identity implies the homomorphism property vector fields fulfilling, 
\begin{equation}
[X_f,X_g]=X_{\{f,g\}} .
\end{equation}

Applying this construction to the generators $x_a$, and restricting to momentum-dependent functions, we obtain vector fields on momentum space,
\begin{equation}
X_{x_a} := (E^{-1})_{a}{}^{\mu}(p)\,\frac{\partial}{\partial p^\mu}.
\label{eq:vectorfield}
\end{equation}
These vector fields are linearly independent and satisfy the same $\mathfrak{su}(1,1)$ algebra as the Poisson generators, now realized as Lie brackets of vector fields on momentum space.

The structures introduced above admit a unified interpretation across classical, geometric, and quantum levels. The Poisson algebra defines the fundamental algebraic structure on phase space. Through the map $f \mapsto X_f$, this algebra is realized geometrically as a Lie algebra of vector fields, while upon quantization it is represented by operator commutators,
\begin{equation}
[\hat f,\hat g] = i\hbar \widehat{\{f,g\}}.
\end{equation}
Thus, the Poisson, Lie, and commutator brackets provide equivalent realizations of the same underlying algebraic structure. In this framework, the classical restriction to functions $F=F(p)$ is mirrored at the quantum level by a choice of polarization.

To construct the quantum theory, it is convenient to introduce a symplectic potential
\begin{equation}
\theta = E_{a\mu}(p)x^a\,dp^\mu ,
\end{equation}
which coincides with the in the first term of the Lagrangian one form: $Ld\tau$ in Eq.~\eqref{260}. The associated symplectic two-form
\begin{equation}
\omega = d\theta
=
E_{a\mu}(p)\,dp^\mu\wedge dx^a
+
x^a\,\partial_\nu E_{a\mu}(p)\,dp^\nu\wedge dp^\mu .
\end{equation}

Closure, $d\omega=0$, (Jacobi Identity) follows from the Maurer--Cartan structure equation
\begin{equation}
\partial_\nu E_{a\mu} - \partial_\mu E_{a\nu}
=
\frac{1}{m_p}\epsilon_{ab}{}^{c} E_{c\nu} E^{b}{}_{\mu}.
\end{equation}

We now proceed to quantization. In geometric quantization, classical observables are first lifted to operators on the full phase space through the Kostant--Souriau prescription, defining a prequantum representation,
\begin{equation}
\hat f=-i\hbar\nabla_{X_f}+f ,
\end{equation}
with connection
\begin{equation}
\nabla=d-\frac{i}{\hbar}\theta
=
d-\frac{i}{\hbar}x^aE_{a\mu}(p)\,dp^\mu .
\end{equation}
At this stage, the wavefunctions depend on both $x^a$ and $p^\mu$, and the representation is not yet irreducible.

Acting with the symplectic potential on the Hamiltonian vector field yields
\begin{align}
\theta(X_{x^a})
&=
x^bE_{b\mu}\,dp^\mu(X_{x^a}) \\
&=
x^bE_{b\mu}(E^{-1}(p))^{a\mu} \\
&=
x^a ,
\end{align}
and therefore the prequantum operator becomes
\begin{equation}
\hat{x}^a_{\mathrm{pre}}
=
-i\hbar (E^{-1}(p))^{a\mu}\partial_\mu
-
\frac{i\hbar}{m_p}\epsilon^{ab}{}_{c}x^c\partial_b .
\end{equation}

To obtain a genuine quantum representation, we impose a polarization. Choosing the momentum polarization
\begin{equation}
\psi(x,p)=\psi(p),
\end{equation}
implies
\begin{equation}
\partial_b\psi(p)=0 .
\end{equation}

In this reduced representation, the second term acts trivially, and the operator descends to
\begin{equation}
\hat{x}^a=-i\hbar (E^{-1}(p))^{a\mu}\partial_\mu .
\end{equation}

Similarly,
\begin{equation}
\hat{p}^\mu=p^\mu .
\end{equation}

These operators reproduce the modified Heisenberg algebra presented in the main text. In this construction, spacetime noncommutativity is encoded in the nontrivial geometry of momentum space, while the choice of polarization selects the physical quantum degrees of freedom.

\section{Computation of the Curvature Scalar}
\addcontentsline{toc}{section}{Appendix: Computation of the Curvature Scalar}

In this appendix, we present two independent but complementary methods to compute the Ricci curvature scalar associated with the curved momentum space geometry that emerges from our framework.

\begin{itemize}
    \item {Method I} utilizes a conformally rescaled dS spacetime metric and derives the curvature scalar using standard conformal transformation formulas.
    \item {Method II} reinterprets the Lie algebra of noncommutative coordinates as the algebra among Killing vector fields defined over a curved momentum space manifold. This algebraic approach employs differential geometric tools and Cartan's structural equation.
\end{itemize}

Both methods yield the same expression for the curvature scalar, thereby confirming the internal consistency of our construction and supporting the geometric interpretation of the underlying noncommutative structure. We now provide the detailed derivation for each method in the subsections that follow.

\subsection{Method I: Conformal Geometry Approach}

Here, we provide a brief derivation of equation~\eqref{ricci}. 
In general, for any pseudo-Riemannian manifold of dimension \(n\), if \(\tilde{R}\) is the Ricci scalar curvature corresponding to the metric \(\tilde{g}_{\mu\nu}\), and \(R\) is the Ricci scalar associated with the conformally related metric \(g_{\mu\nu}\), connected via the conformal factor \(\Omega^2\), then the two curvatures are related by the following formula:

\begin{align}
R_{\mu\nu} &= \tilde{R}_{\mu\nu} 
- \tilde{\nabla}_\mu \tilde{\nabla}_\nu \ln \Omega 
- \tilde{g}_{\mu\nu} \tilde{\Box} \ln \Omega 
+ \tilde{\nabla}_\mu \ln \Omega \, \tilde{\nabla}_\nu \ln \Omega 
- \tilde{g}_{\mu\nu} \tilde{\nabla}^\lambda \ln \Omega \, \tilde{\nabla}_\lambda \ln \Omega \\
R &= \frac{1}{\Omega^2} \left[ \tilde{R} 
- 2(n-1) \tilde{\Box} \ln \Omega 
- (n-1)(n-2) \, \tilde{\nabla}^\lambda \ln \Omega \, \tilde{\nabla}_\lambda \ln \Omega \right]
\end{align}

Here, \(\tilde{\Box}\) and \(\tilde{\nabla}\) denote the d'Alembertian and the covariant derivative operators compatible with the metric \(\tilde{g}_{\mu\nu}\), respectively.

\vspace{0.2cm}

\noindent
Now, the explicit form of the d'Alembertian acting on \(\ln \Omega\) is given by:

\begin{align}
    \tilde{\Box} \ln \Omega &= \tilde{\nabla}_{\mu} \tilde{\nabla}^{\mu} \ln \Omega \\
    &= \frac{1}{\sqrt{|\tilde{g}|}} \, \partial_{\mu} \left( \sqrt{|\tilde{g}|} \, \tilde{g}^{\mu\nu} \, \partial_{\nu} \ln \Omega \right)
\end{align}

where the inverse metric and its determinant are given by:

\begin{align}
    \tilde{g}^{\mu\nu} &= \eta^{\mu\nu} -m_{p}^{\hspace{4pt}-2} p^{\mu}p^{\nu} \label{inv} \\
    \tilde{|g|} &= \frac{1}{1 - m_{p}^{\hspace{4pt}-2} p^2} \label{d}
\end{align}

Using the conformal factor from equation~\eqref{confo}, along with equations~\eqref{inv} and~\eqref{d}, we find:

\begin{align}
    \tilde{\Box} \ln \Omega &= 3m_{p}^{\hspace{4pt}-2} + \frac{m_{p}^{-4}p^2}{1 -m_{p}^{-2} p^2}
\end{align}

Furthermore, the square of the gradient of \(\ln \Omega\) evaluates to:

\begin{align}
    \left( \tilde{\nabla} \ln \Omega \right)^2 &= \tilde{g}^{\mu\nu} \, \partial_{\mu} \ln \Omega \, \partial_{\nu} \ln \Omega \\
    &= \left( \eta^{\mu\nu} -m_{p}^{-2} p^{\mu}p^{\nu} \right) \frac{m_{p}^{-2} p_{\mu}}{1 -m_{p}^{-2} p^2} \frac{m_{p}^{-2} p_{\nu}}{1 - m_{p}^{-2}p^2} \\
    &= \frac{m_{p}^{-4}p^2}{1 - m_{p}^{-2} p^2}
\end{align}

Finally, for \(n = 3\), which corresponds to our case, substituting these results into the general expression gives:

\begin{align} \label{ricci}
    R_{\mu\nu} &= -\frac{1}{m_p^2} g_{\mu\nu} \\
    R &= \tilde{R}(1 - \frac{p^2}{m^{2}_{p}}) + \frac{6p^2}{m_p^4} - \frac{12}{m_p^2}
\end{align}

\subsection{Method II: Killing Vector Fields Approach}

In this approach, we reinterpret the noncommutative coordinate algebra as the algebra of Killing vector fields \(\{\hat{x}_a\}\) defined on a curved momentum space. The underlying metric structure (\ref{21}) of this manifold has been established in the main text, and its curvature is entirely governed by the associated Killing algebra.

These Killing vector fields obey the algebra
\begin{equation}
    [x_a, x_b] = C_{ab}^{~~c} x_c, \qquad \text{with} \quad C_{ab}^{~~c} = 2\lambda \epsilon_{ab}^{~~c}.
\end{equation}
The components of the Killing vector fields in the tangent space \( T_Q(\mathcal{P}_{m_p}) \) are given by
\begin{equation}
    x_a := \left(E^{-1}(p)\right)_a{}^{\mu} \frac{\partial}{\partial p^{\mu}}.
    \label{15}
\end{equation}
The algebra of Killing vector fields in the same tangent space \( T_Q(\mathcal{P}_{m_p}) \) can be expressed in terms of their components as follows:
\begin{equation}
    [x_a, x_b] = \left([x_a, x_b]^{\mu}\right) \frac{\partial}{\partial p^{\mu}}.
\end{equation}

Therefore, by employing the general framework of differential geometry on curved momentum space manifolds, and noting that the curvature is fully determined by the Killing algebra, the Ricci curvature tensor can be written in the form
\begin{equation}
    R_{~\mu}^{\delta} = E^a{}_{\mu}\left[\nabla_{\nu},\nabla^{\delta} \right](E^{-1})_a{}^{\nu}.
\end{equation}
Since $\lbrace\hat{x}_a \rbrace $ are Killing vector fields,
\begin{equation}
    \nabla_{\nu}(E^{-1})_a{}^{\nu} = 0.
\end{equation}
As such, we can rewrite the Ricci curvature tensor as
\begin{equation}
\begin{aligned}
    R_{~\mu}^{\delta} &= E^a{}_{\mu} \nabla_{\nu} \nabla^{\delta} (E^{-1})_a{}^{\nu}\\
    &= \frac{1}{2} E^a{}_{\mu} \nabla_{\nu} \mathrm{F}_a{}^{\delta\nu}
\end{aligned}
\end{equation}
where \(\frac{1}{2} \mathrm{F}_a{}^{\mu\nu} = \nabla^{[\mu} (E^{-1})_a{}^{\nu]}\).

The Ricci curvature tensor can further be written as
\begin{equation}
    R_{~\mu}^{\delta} = \frac{1}{2} \nabla_{\nu} \left( E^a{}_{\mu} \mathrm{F}_a{}^{\delta\nu} \right) + \frac{1}{4} \mathrm{F}^a{}_{\mu\nu} \mathrm{F}_a{}^{\delta\nu}.
\end{equation}
Here we used the Killing condition to get the last term. Now we evaluate the first term using the Cartan structural equation:
\begin{equation} \label{structCEq}
    \mathrm{F}_a{}^{\delta\nu} = C_a{}^{bc} (E^{-1})_b{}^{\nu} (E^{-1})_c{}^{\delta}.
\end{equation}
Then,
\begin{equation}
\begin{aligned}
    \nabla_{\nu} \left( E^a{}_{\mu} \mathrm{F}_a{}^{\delta\nu} \right) &= \nabla_{\nu} \left( E^a{}_{\mu} C_a{}^{bc} (E^{-1})_b{}^{\nu} (E^{-1})_c{}^{\delta} \right)\\
    &= C_a{}^{bc} (E^{-1})_b{}^{\nu} \nabla_{\nu} \left( E^a{}_{\mu} (E^{-1})_c{}^{\delta} \right)\\
    &= -C^{bac} (E^{-1})_b{}^{\nu} \nabla_{\nu} \left( E_{a\mu} (E^{-1})_c{}^{\delta} \right)
\end{aligned}
\end{equation}
In the second line we used the constancy of structure constants and the Killing condition. In the third line, total antisymmetry of \(C^{abc}\) is used.

Expanding the derivative:
\begin{equation}
\begin{aligned}
    \nabla_{\nu} \left( E^a{}_{\mu} \mathrm{F}_a{}^{\delta\nu} \right)
    &= -C^{bac} C_{ade} (E^{-1})_b{}^{\nu} (E^{-1})_e{}^{\nu} E^d{}_{\mu} (E^{-1})_c{}^{\delta} \\
    &\quad - C^{bac} C_{cde} (E^{-1})_b{}^{\nu} (E^{-1})_e{}^{\nu} E_{a\mu} (E^{-1})_d{}^{\delta}
\end{aligned}
\end{equation}

Relabeling dummy indices:
\begin{equation}
\begin{aligned}
    \nabla_{\nu} \left( E^a{}_{\mu} \mathrm{F}_a{}^{\delta\nu} \right)
    &= -C^{bac} C_{adb} E^d{}_{\mu} (E^{-1})_c{}^{\delta} 
    - C_{d}{}^{ba} C^{c}{}_{ae} E^d{}_{\mu} (E^{-1})_c{}^{\delta} \\
    &= -C^{bac} C_{adb} E^d{}_{\mu} (E^{-1})_c{}^{\delta}
    + C_{bad} C^{acb} E^d{}_{\mu} (E^{-1})_c{}^{\delta} \\
    &= 0
\end{aligned}
\end{equation}

Hence, the Ricci tensor becomes
\begin{equation}
    R_{~\mu}^{\delta} = \frac{1}{4} \mathrm{F}^a{}_{\mu\nu} \mathrm{F}_a{}^{\delta\nu}.
\end{equation}

From \eqref{structCEq} it follows that
\begin{equation}
    R_{\mu\nu} = -2\lambda^2 g_{\mu\nu}
\end{equation}
and finally, the curvature scalar is
\begin{equation}
    R = g^{\mu\nu} R_{\mu\nu} = -\frac{6}{m^{2}_{p}},
\end{equation}
where we have substituted \(\lambda = \frac{1}{m_p}\). This result is consistent with our earlier determination that the momentum space has the geometry of an \(\mathrm{AdS}_3\) manifold.

\section{Deriving the Configuration-Space Action with Curved Momentum Space Corrections}

We begin with the first-order action for a relativistic point particle whose dynamics are modified by a curved momentum space geometry:
\begin{equation}
S[q, p, \Lambda] = \int d\tau \left( \dot{q}^\mu p_\mu - \Lambda \left[ f(p^2) + M^2 \right] \right),
\end{equation}
where the deformation is encoded in the function
\begin{equation}
f(p^2) = -m_p^2 \left[ \tan^{-1} \left( \frac{\sqrt{-p^2}}{m_p} \right) \right]^2.
\end{equation}

Varying the action with respect to \( p^\mu \) yields
\begin{equation}
p^\mu = \frac{1}{2\Lambda} \left( \frac{\partial f}{\partial p^2} \right)^{-1} \dot{q}^\mu.
\label{eq:p-mu}
\end{equation}
We compute the derivative as
\begin{equation}
\frac{\partial f}{\partial p^2} = \frac{m_p \tan^{-1}\left( \frac{\sqrt{-p^2}}{m_p} \right)}{ \sqrt{-p^2} \left( 1 + \left( \frac{\sqrt{-p^2}}{m_p} \right)^2 \right)}.
\end{equation}

Let us define the dimensionless variable \( u := \frac{\sqrt{-p^2}}{m_p} \), so that \( p^2 = -m_p^2 u^2 \). Then the inverse derivative becomes
\begin{equation}
\left( \frac{\partial f}{\partial p^2} \right)^{-1} = \frac{u(1 + u^2)}{\tan^{-1}(u)}.
\end{equation}

From the definition of \( p^\mu \), we find the square of the momentum:
\begin{equation}
p^2 = \frac{\dot{q}^2}{4\Lambda^2} \left( \frac{u(1 + u^2)}{\tan^{-1}(u)} \right)^2 = -m_p^2 u^2, \qquad \text{with } \dot{q}^2 = \dot{q}^\mu \dot{q}_\mu.
\end{equation}

Solving this relation perturbatively in the regime \( |p^2| \ll m_p^2 \), we obtain
\begin{equation}
\frac{\sqrt{-\dot{q}^2}}{2\Lambda m_p} = u - \frac{4}{3} u^3 + \mathcal{O}(u^5).
\end{equation}
 Inverting this expression to express \(\sqrt{-p^2} \) in terms of \( \dot{q}^\mu \), we find
\begin{equation}
\frac{\sqrt{-p^2}}{m_p}
=
\frac{\sqrt{-\dot{q}^2}}{2 \Lambda m_p}
+
\frac{1}{6} \cdot \frac{\left(-\dot{q}^2\right)^{3/2}}{\Lambda^3 m_p^3}
+
\mathcal{O}(m_p^{-5})
\end{equation}

Substituting this into Eq.~\eqref{eq:p-mu}, we obtain the leading-order expression for the momentum:
\begin{equation}
p^{\mu} = \frac{1}{2\Lambda} \left[ 1 - \frac{\dot{q}^2}{3 \Lambda^2 m_p^2} + \mathcal{O} \left( \frac{\dot{q}^4}{\Lambda^4 m_p^4} \right) \right] \dot{q}^{\mu}.
\end{equation}

Expanding \( f(p^2) \) in powers of \( u \), we find
\begin{align}
f(p^2) &= -m_p^2 \left[ \tan^{-1}(u) \right]^2 \\
&= -m_p^2 \left[ u - \frac{u^3}{3} + \cdots \right]^2 \\
&= \frac{\dot{q}^2}{4\Lambda^2} - \frac{(\dot{q}^2)^2}{24 \Lambda^4 m_p^2} + \mathcal{O} \left( \frac{(\dot{q}^2)^3}{\Lambda^6 m_p^4} \right).
\end{align}

With these results, the first-order action becomes
\begin{equation}
S[q, \Lambda] = \int d\tau \left[
\frac{\dot{q}^2}{4\Lambda}
- \frac{(\dot{q}^2)^2}{24 \Lambda^3 m_p^2}
- \Lambda M^2
\right].
\end{equation}

Varying with respect to \( \Lambda \), and solving to leading order in \( 1/m_p^2 \), we obtain:
\begin{equation}
\Lambda = \sqrt{ -\frac{\dot{q}^2}{4M^2} } \left( 1 + \frac{M^2}{m_p^2} + \mathcal{O} \left( \frac{1}{m_p^4} \right) \right).
\end{equation}

Substituting this result back into the action gives the effective configuration-space Lagrangian, now fully expressed in terms of \( \dot{q}^\mu \), \( M \), and \( m_p \), consistently up to \( \mathcal{O}(1/m_p^2) \):
\begin{equation}
S_{eff}[q] = \int d\tau\, \mathcal{L}_{\text{eff}} = \int d\tau \left[
- M \left( 1 + \frac{M^2}{3 m_p^2} \right) \sqrt{-\dot{q}^2}
- \frac{M^3}{3 m_p^2} (-\dot{q}^2)^{5/2}
+ \mathcal{O} \left( \frac{1}{m_p^4} \right)
\right].
\end{equation}

The first term captures a renormalization of the inertial mass induced by the curvature of momentum space. The second term encodes a higher-order velocity-dependent contribution that arises within this perturbative elimination of the auxiliary variables. Its detailed form is sensitive to the perturbative scheme and parametrization adopted in reducing the first-order formulation, and should therefore be regarded as an effective contribution rather than a unique leading-order prediction. The dominant physical effect at this order is governed by the renormalized mass parameter.


\section{Worldline Dynamics and Geodesic Consistency}

In this appendix, we analyze the dynamics of a particle governed by a non-standard worldline action and demonstrate that, despite the modified Lagrangian, the resulting trajectories remain geodesic in a fixed background spacetime.

We consider a class of effective worldline actions of the form

\begin{equation}
S[x] = \int d\tau\, L(X),
\qquad
X = - g_{\mu\nu}(x)\dot{x}^\mu \dot{x}^\nu ,
\end{equation}
with a representative Lagrangian of the form
\begin{equation}
L(X) = -\alpha X^{1/2} - \beta X^{5/2}.
\end{equation}

Such forms arise as effective configuration-space representations obtained after eliminating auxiliary variables from a reparametrization-invariant first-order formulation.

Here $\tau$ is an arbitrary parameter along the worldline, and $X$ corresponds to the squared norm of the three-velocity. Since $L$ is not homogeneous of degree one in $\dot{x}^\mu$, this representation is not manifestly invariant under arbitrary reparametrizations of $\tau$ at the off-shell level. This leads to equations of motion that are, in general, non-affine.

Varying the action with respect to $x^\mu$ yields the Euler--Lagrange equations,
\begin{equation}
\frac{d}{d\tau}\left(\frac{\partial L}{\partial \dot{x}^\rho}\right)
-
\frac{\partial L}{\partial x^\rho}
=0.
\end{equation}
Using the fact that $L=L(X)$, together with
\begin{equation}
\frac{\partial X}{\partial \dot{x}^\rho} = -2 g_{\rho\nu}\dot{x}^\nu,
\end{equation}
and taking into account the spacetime dependence of the metric, one arrives after straightforward algebra at
\begin{equation}
\ddot{x}^\rho
+
\Gamma^\rho_{\alpha\beta}\dot{x}^\alpha \dot{x}^\beta
=
-\frac{\dot{L}_X}{L_X}\dot{x}^\rho,
\end{equation}
where $L_X \equiv dL/dX$, and $\Gamma^\rho_{\alpha\beta}$ are the Christoffel symbols of the background metric.

Introducing the three-velocity $u^\mu = \dot{x}^\mu$, this equation can be written in covariant form as
\begin{equation}
D_\tau u^\mu = \lambda(\tau)\,u^\mu,
\qquad
\lambda(\tau) = -\frac{\dot{L}_X}{L_X}.
\end{equation}
At this stage, the motion appears non-affine, with $\lambda(\tau)$ playing the role of a parameter-dependent acceleration along the trajectory.

To determine whether this non-affinity is physical or merely an artifact of parametrization, we examine the evolution of the velocity norm. Taking the derivative of $u^2 = g_{\mu\nu}u^\mu u^\nu$, we obtain
\begin{equation}
\frac{d}{d\tau}(u^2)
=
2 g_{\mu\nu} u^\mu D_\tau u^\nu
=
2\lambda\,u^2.
\end{equation}
Using $X = -u^2$, this implies
\begin{equation}
\dot{X} = -2\lambda X.
\end{equation}
Substituting $\lambda = -\dot{L}_X/L_X$ and using $\dot{L}_X = L_{XX}\dot{X}$, we find
\begin{equation}
\dot{X}
\left(
1 - 2\frac{L_{XX}}{L_X}X
\right)
= 0.
\end{equation}

This equation admits two possible branches. The first corresponds to $\dot{X}=0$, while the second would require
\begin{equation}
1 - 2\frac{L_{XX}}{L_X}X = 0.
\end{equation}
For the specific Lagrangian under consideration, the second condition is not satisfied identically along the trajectory and therefore does not define a consistent dynamical branch. Consequently, the only physically admissible solution is
\begin{equation}
\dot{X} = 0,
\end{equation}
which implies that $X$ is conserved along the worldline.

It follows immediately that $\dot{L}_X = 0$, and hence $\lambda(\tau)=0$. The equation of motion then reduces to
\begin{equation}
D_\tau u^\mu = 0,
\end{equation}
which is precisely the affine geodesic equation associated with the Levi--Civita connection of the background metric.

We thus conclude that, despite the non-standard form of this effective worldline action and the absence of manifest reparametrization invariance off shell, the physical trajectories correspond to affine geodesics. In particular, the motion is independent of the particle's internal parameters $\alpha$ and $\beta$, and depends only on the background geometry.

Accordingly, at the level of spacetime trajectories in a fixed background, the weak equivalence principle—understood as the universality of free fall—remains preserved. The effects of curved momentum space do not modify the geodesic structure itself but instead appear in the kinematical and dynamical relations, namely in the modified mass-shell (dispersion) relation, the relation between momentum and velocity, and the effective stress-energy tensor sourcing gravity.

\section{Holonomy Analysis}

Holonomy computations are useful because they capture the global
properties of the spacetime. In $(2+1)$–dimensional gravity the mass
content of the geometry is encoded in the holonomy of the connection
around the origin. In particular, the deficit angle represents a
topological invariant that determines how different patches of
$AdS_3$ are glued together at $r=0$.

\vspace{0.2cm}
\noindent 
Following \cite{Banados1994,Ammon2013,Castro:2020smu}, we compute the
holonomy for the two ranges of the dimensionless parameter $\mu$:
(i) $\mu\in[-1,0)$ corresponding to the conical defect (particle)
regime and (ii) $\mu\ge0$ corresponding to the BTZ black hole regime.

\vspace{0.2cm}
\noindent 
Let us begin by re-writing ~(\ref{169}) 
\begin{equation}
ds^2 = -f^2(r)\,dt^2 + \frac{1}{f^2(r)}\,dr^2 + r^2 d\phi^2;
\label{B1}\quad 
f^2(r)=\frac{r^2}{l^2}-\mu
\end{equation}
The full set of dreibeins, that can just be read-off from here are
\begin{equation}
e^0 = f(r)\,dt, \qquad
e^1 = \frac{1}{f(r)}\,dr, \qquad
e^2 = r\,d\phi
\label{B2}
\end{equation}

which fulfils,
\begin{equation}
   ds^2 = \eta_{ab}\,e^a e^b 
\end{equation}

with $a,b \in [0,1,2]$. They provide the orthonormal basis for the cotangent space defined at each point of $AdS_3$ $(r>0)$. Since we are interested in carrying out a closed loop integration, where both $r$ and $t$ are held fixed, it will be enough to obtain the $\phi$-component $A_\phi$ of the connection one-form that is required to be integrated.

\vspace{0.2cm}
\noindent 
Now given that the isometry group of $AdS_3$ is
\[
SO(2,2)\simeq SL(2,\mathbb{R})\times SL(2,\mathbb{R}),
\]
the gravitational variables can be organized into two independent
$sl(2,\mathbb{R})$--valued gauge connections. After gauging this
symmetry one obtains
\begin{equation}
A^{(\pm)} = A^{(\pm)a}_\mu J_a\, dx^\mu ,
\end{equation}
with
\begin{equation}
A^{(\pm)a} = \omega^a \pm \frac{1}{\ell} e^a ,
\label{B3}
\end{equation}
where $e^a$ denotes the dreibein and $\omega^a$ is the spin connection
one--form. The latter is defined through the dualization
\begin{equation}
\omega^a = \frac{1}{2}\epsilon^{a}{}_{bc}\,\omega^{bc}_\mu\, dx^\mu .
\end{equation}

Here $J_a$ are the generators of the $sl(2,\mathbb{R})$ algebra.
In the fundamental $(2\times2)$ representation they can be written
in terms of the Pauli matrices as
\begin{equation}
J_0=-\frac{i}{2}\sigma_2, \qquad
J_1=\frac{1}{2}\sigma_3, \qquad
J_2=\frac{1}{2}\sigma_1 ,
\end{equation}
which satisfy the algebra
\[
[J_a,J_b]=\epsilon_{ab}{}^{c}J_c .
\]
With our conventions the raised generators satisfy
$J^0=-J_0$, $J^1=J_1$ and $J^2=J_2$.

As shown by Witten \cite{Witten:1988hc}, Einstein gravity in
$(2+1)$ dimensions with negative cosmological constant can be
formulated equivalently as the difference of two Chern--Simons actions,
\[
S_{CS}[A^{(+)}]-S_{CS}[A^{(-)}],
\]
where
\begin{equation}
S_{CS}(A^{(\pm)})=
\frac{\ell}{16\pi G}
\int
\mathrm{Tr}
\left(
A^{(\pm)}\wedge dA^{(\pm)}
+\frac{2}{3}A^{(\pm)}\wedge A^{(\pm)}\wedge A^{(\pm)}
\right).
\end{equation}

The equations of motion obtained by varying $A^{(\pm)}$ imply that the
corresponding field strengths are flat,
\begin{equation}
F^{(\pm)}=dA^{(\pm)}+A^{(\pm)}\wedge A^{(\pm)}=0 ,
\end{equation}
everywhere except at the location of the point source. In the presence
of a particle located at $r=0$, the curvature acquires a distributional
contribution of the form
\begin{equation}
F^{(\pm)}
=
2\pi\sqrt{-\mu}\,
\delta^{(2)}(x)\,
dx^1\wedge dx^2 ,
\end{equation}
which encodes the conical defect produced by the mass parameter $\mu$
in the regime $\mu<0$.
Now the holonomy computation can be carried out by using any one of this pair of connections $A^{(+)}$ or $A^{(-)}$ in our case of zero angular momentum sector. Therefore, we choose, without loss of generality, to work with $A^{(+)}$ and which henceforth will simply be denoted by $A$, by suppressing the superscript $'+'$.

\vspace{0.2cm}
\noindent 
To begin with, let us determine the spin connection one-form $\omega^a$ using Cartan's equation of structure (assuming torsion $=0$)

\begin{equation}
de^a + \omega^a_{\ b} \wedge e^b = 0
\label{B4}
\end{equation}

A straightforward computation for the $a=0$ component yields
\begin{equation}
\omega^0=\omega^{1}_{\ 2}=-f(r)\,d\phi .
\label{B5}
\end{equation}

Although the radial component $A_r\neq0$, the holonomy around a
circular loop depends only on the $\phi$–component of the
$sl(2,\mathbb{R})$ Chern–Simons connection. This component in the fundamental $2\times2$ representation  is given by, 

\begin{equation}
A_\phi=\omega^{12}_{\ \phi}J_0+\frac{1}{\ell}e^{2}_{\ \phi}J_2
      =-\sqrt{\frac{r^2}{\ell^2}-\mu}\,J_0+\frac{r}{\ell}J_2
      =\frac{1}{2}
\begin{pmatrix}
0 & \left(\frac{r}{\ell}+f(r)\right)\\
\left(\frac{r}{\ell}-f(r)\right) & 0
\end{pmatrix}.
\label{B6}
\end{equation}
where, $J_0=-\frac{i}{2}\sigma_2,
J_1=\frac{1}{2}\sigma_3,
J_2=\frac{1}{2}\sigma_1$ 
At this stage one finds
\[
\mathrm{Tr}(A_{\phi}^2)=\frac{\mu}{2},
\]
showing that the holonomy invariant depends only on the mass parameter
$\mu$. Since the Chern–Simons connection is flat ($F=0$) everywhere
away from the source at $r=0$, the global information about the
geometry is encoded entirely in the holonomy.

The explicit $r$–dependence of $A_\phi$ can be removed by performing
an $SL(2,\mathbb{R})$ gauge transformation \cite{Coussaert1995}

\begin{equation}
A \rightarrow g^{-1}Ag + g^{-1}dg ,
\qquad g\in SL(2,\mathbb{R}).
\label{B7}
\end{equation}

Choosing the radial gauge transformation

\begin{equation}
g(r)=
\exp\!\left[
\ln\!\left(
\frac{r}{\ell}+\sqrt{\frac{r^2}{\ell^2}-\mu}
\right)
\frac{J_0}{\sqrt{-\mu}}
\right],
\label{B8}
\end{equation}

aligns the local frame with the global symmetry directions.  In the
asymptotic region $r\to\infty$ this transformation reduces to

\begin{equation}
g(r)\rightarrow b(r)=\exp\!\left[\ln\!\left(\frac{r}{\ell}\right)J_0\right].\label{aaa}
\end{equation}

In this radial gauge the connection becomes

\begin{equation}
A_\phi=\sqrt{-\mu}\,J_0,
\qquad
A_r=0.
\label{B9}
\end{equation}

The holonomy around a closed loop $\gamma$ encircling the origin is then
defined by

\begin{equation}
W[\gamma]=\mathcal{P}\exp\oint_\gamma A .
\label{B10}
\end{equation}

For a circular loop with winding number one, this simplifies to

\begin{equation}
W[\gamma]=e^{2\pi\sqrt{-\mu}\,J_0}.
\label{B11}
\end{equation}

In the fundamental $(2\times2)$ representation this becomes

\begin{equation}
W[\gamma]=
\begin{pmatrix}
\cos(\pi\sqrt{-\mu}) & -\sin(\pi\sqrt{-\mu})\\
\sin(\pi\sqrt{-\mu}) & \cos(\pi\sqrt{-\mu})
\end{pmatrix}.
\label{B12}
\end{equation}

The corresponding Wilson loop is

\begin{equation}
\mathrm{tr}\,W[\gamma]=2\cos(\pi\sqrt{-\mu})<2 .
\label{B13}
\end{equation}

This characterizes the elliptic holonomy associated with the conical
defect geometry describing a point particle with mass parameter
$\mu\in[-1,0)$.

For completeness we also quote the holonomy in the adjoint
$(3\times3)$ representation, where the generators are

\[
(J_a)^b_{\ c}=\epsilon_{ac}^{\ \ b}.
\]

In this representation

\begin{equation}
W_{\text{adj}}[\gamma]=
\begin{pmatrix}
\cos(2\pi\sqrt{-\mu}) & \sin(2\pi\sqrt{-\mu}) & 0\\
-\sin(2\pi\sqrt{-\mu}) & \cos(2\pi\sqrt{-\mu}) & 0\\
0 & 0 & 1
\end{pmatrix}.
\label{B14}
\end{equation}

This form makes the geometrical interpretation transparent.  The
holonomy corresponds to a rotation by angle $2\pi\sqrt{-\mu}$,
allowing the deficit angle to be identified as

\[
\Delta\phi=2\pi(1-\sqrt{-\mu}),
\]

which reproduces the earlier result.

Finally, in the BTZ regime $\mu>0$ the quantity $\sqrt{-\mu}$ becomes
imaginary. This reflects the change of topology associated with the
appearance of an event horizon at

\[
r_+=\ell\sqrt{\mu}.
\]

After performing the appropriate
radial gauge transformation by using one of the boost generators $J_1$ or $J_2$ i.e. taking $b(r)=\exp\!\left[\ln\!\left(\frac{r}{\ell}\right)J_1\right]$ as in ~\cite{Banados1994} one finds, 

\begin{equation}
a_\phi=\sqrt{\mu}
\begin{pmatrix}
0 & 1\\
1 & 0
\end{pmatrix}.
\label{B16}
\end{equation}

Exponentiating the connection gives

\begin{equation}
W[\gamma]=e^{2\pi a_\phi}=
\begin{pmatrix}
\cosh(\pi\sqrt{\mu}) & \sinh(\pi\sqrt{\mu})\\
\sinh(\pi\sqrt{\mu}) & \cosh(\pi\sqrt{\mu})
\end{pmatrix}
\in SO(1,1).
\label{B17}
\end{equation}

The Wilson loop therefore becomes

\begin{equation}
\mathrm{Tr}\,W[\gamma]=2\cosh(\pi\sqrt{\mu})\ge2 .
\end{equation}

The equality holds only for $\mu=0$, corresponding to the
massless BTZ solution. For $\mu>0$ the holonomy is hyperbolic,
characteristic of the BTZ black hole geometry.

The mass parameter determined from the holonomy is therefore

\begin{equation}
M_{\text{ADM}}=\frac{\mu}{8G}\ge0 .
\end{equation}

\section{Affine Parameter for the radial null geodesics for BTZ geometry}
In subsection 5.3 we had computed the return coordinate time of a massless particle, taken as a probe, to travel back and forth from the horizon to $AdS_3$ boundary and it turned out be a finite quantity. Here however we would like to show that an affine parameter can also be used to parametrize such a null trajectory, which however turns out be divergent.\\

Let '$s$' be such an affine parameter. Now using
\begin{eqnarray}\label{96}
    \dot{x}^{\mu} &=& \frac{dx^{\mu}}{ds}, 
    \end{eqnarray}
We can now make use of \eqref{1460} to write,
\begin{eqnarray}
    \dot{t}=\pm\frac{1}{f^2(r)}\dot{r}
\end{eqnarray}

At this stage we observe, $\vec{V}=v^{\mu}\partial_{\mu}=\partial_{t}$ is a global timelike Killing vector for BTZ metric.\eqref{metric}, where the component are identified as, 
\begin{eqnarray}
    v^{0}=1,v^{i}=0 \quad (i=1,2)
\end{eqnarray}
The corresponding conserved quantity is energy 
\begin{eqnarray}
    E=\left|\int d^2x\sqrt-g T^{0\mu}(x)v_{\mu}\right|
\end{eqnarray}
where
\begin{eqnarray}
    T^{\mu\nu}(x)=\frac{1}{\sqrt{-g}}\int p^{\mu}dx^{\nu} \delta^3(x-x(s)) \quad ; x^0(s)=s
\end{eqnarray} is the EM tensor of a massless particle, moving along the above null geodesic. 
Here the integration is taken along the null trajectory.In component form 
\begin{eqnarray}
    T^{0\mu}=\frac{p^{\mu}} {\sqrt{-g}}\delta^2(\vec{x}-\vec{x}(s))
\end{eqnarray}represents density of three momentum.With this one gets, 
\begin{eqnarray}
    E = \left| \int d^2x \, \sqrt{-g}\, T^{0\mu} v_{\mu} \right|
   = \left| \int d^2x \, \sqrt{-g}\, T^{00} g_{00} \right|
\end{eqnarray}
Here we have made use of the digonal form of the metric tensor $g_{\mu\nu}$\eqref{72}. Now identifying, $p^{\mu}=\frac{dx^{\mu}}{ds}=\dot{x^{\mu}}$, by choosing a suitable scale for "$s$", we get the conserved energy $E$ to be given by, 
\begin{eqnarray}
    E=\left|\int d^2x\dot{x}^0g_{00}(x)\delta^2(\vec{x}-\vec{x}(s))\right|=f^2(r)\dot{t}=\dot{r} 
\end{eqnarray}
We can now  make use of \eqref{96} to identify, 
\begin{eqnarray}
    r=Es
\end{eqnarray}
This implies that $r\rightarrow \infty$ as $s\rightarrow\infty$, which is indicative of the geodesically complete nature of the manifold. But the important point to note that the time taken as, measured by the affine parameter s, for the massless particle to escape to infinity indeed diverges- unlike that of the BTZ coordinate time as measured by an observer located either at the $AdS_3$ boundary or near the horizon. \\

\end{document}

%% file: mathsym.tex
\newcommand{\pb}[2]{\left\{ \hat{#1}, \hat{#2} \right\}}

\newcommand{\kd}[2]{{\delta_{#1}}^{#2}}
\newcommand{\ikd}[2]{{\delta^{#1}}_{#2}}

\newcommand{\eps}[3]{{{\epsilon^{#1}}_{#2#3}}}
\newcommand{\com}[2]{\left[#1,#2\right]}

\newcommand{\epc}[3]{{\epsilon^{#1#2#3}}}

\newcommand{\vier}[3]{  { {\left(E^{-1}\left(#3\right)\right)}^{#1}}_{#2}}

\newcommand{\exvielm}[3]{{\left(E^{-1}\left({#1}\right)\right)^{#2}}_{#3}}

\newcommand{\exviecm}[3]{{\left(E^{-1}\left({#1}\right)\right)_{#2}}^{#3}}

\newcommand{\hcut}{\hbar}
\newcommand{\trib}[2]{{{(E^{-1})}_{#1}}^{#2}}
\newcommand{\tribform}[5]{ {\delta_{#1}}^{#2} +  #4 {{\epsilon_{#1}}^{#2#3}}  p_{#3} - {#5} p_{#1}p^{#2} }

\newcommand{\pd}[2]{\frac{\partial#1}{\partial#2}}